\documentclass[10pt,english]{IEEEtran}

\usepackage[english]{babel}
\usepackage{amssymb, amsmath}
\usepackage{graphicx}
\usepackage{subfigure}
\usepackage{hyperref}
\usepackage{multirow}
\usepackage{enumitem}
\usepackage{dblfloatfix}
\usepackage[ruled,vlined,linesnumbered]{algorithm2e}

\newcommand{\norm}[2]{\| #1 \|_{#2}}
\newcommand{\abs}[1]{| #1 |}
\newcommand{\expval}[1]{\mathbb{E} \left [ #1 \right]}
\newcommand{\argmin}{\mathrm{arg \, min}}
\newcommand{\argmax}{\mathrm{arg \, max}}
\newcommand{\diag}[1]{\mathrm{diag} \left ( #1 \right)}
\newcommand{\prob}[1]{\mathrm{Pr} \left ( #1 \right)}
\newcommand{\cscgdist}[2]{\sim \mathcal{CN} \left( #1, #2 \right)}

\newcommand{\complexset}{\mathbb{C}}
\newcommand{\realset}{\mathbb{R}}
\newcommand{\integerset}{\mathbb{Z}}
\newcommand{\set}[1]{\left\{ #1 \right\}}

\SetAlFnt{\small}
\SetAlCapFnt{\small}

\begin{document}

\title{QoS-Aware User Scheduling in Crowded XL-MIMO Systems Under Non-Stationary Multi-State LoS/NLoS Channels}

\author{
{Jo\~ao Henrique Inacio de Souza}, 
{José Carlos Marinello}, 
{Abolfazl Amiri}, 
{Taufik Abr\~ao}
\thanks{This work was supported in part by the National Council for Scientific and Technological Development (CNPq) of Brazil under Grant 310681/2019-7 and (Scholarship)}

\thanks{J. H. I. de Souza and T. Abr\~ao are with the Electrical Engineering Department, State University of Londrina, PR, Brazil.  E-mail: {\scriptsize joaohis@outlook.com; \scriptsize taufik@uel.br}.}
\thanks{J. C. Marinello is with Electrical Engineering Department, Federal University of Technology PR, Cornélio Procópio, PR, Brazil.  E-mail: {\scriptsize jcmarinello@utfpr.edu.br}.}
\thanks{A.  Amiri, is with the Department of Electronic Systems,  Technical Faculty of IT and Design; Aalborg University,	Denmark; E-mail: {\scriptsize aba@es.aau.dk}}
}

\maketitle

\vspace{7mm}

\begin{abstract}
Providing minimum quality-of-service (QoS) in crowded wireless communications systems, with high user density, is challenging due to the network structure with limited transmit power budget and resource blocks. Smart resource allocation methods, such as user scheduling, power allocation, and modulation and coding scheme selection, must be implemented to cope with the challenge.
Aiming to enhance the number of served users with minimum QoS in the downlink (DL) channel of crowded extra-large scale massive multiple-input multiple-output (XL-MIMO) systems, in this paper we propose a QoS-aware joint user scheduling and power allocation technique. The proposed technique is constituted by two sequential procedures: the clique search-based scheduling (CBS) algorithm for user scheduling followed by optimal power allocation with transmit power budget and minimum achievable rate per user constraints.
To accurately evaluate the proposed technique in the XL-MIMO scenario, we propose a generalized non-stationary multi-state channel model based on spherical-wave propagation assuming that users under LoS and NLoS transmission coexist in the same communication cell. Such model considers that users under different channel states experience different propagation aspects both in the multi-path fading model and the path loss rule.
Numerical results on the achievable sum-rate, number of scheduled users, and distribution of the scheduled users reveal that the proposed CBS algorithm provides a fair coverage over the whole cell area, achieving remarkable numbers of scheduled users when users under the LoS and NLoS channel states coexist in the communication cell.
\end{abstract}


\section{Introduction}\label{sec:introduction}

Extra-large scale massive multiple-input multiple-output (XL-MIMO) are the deployments of massive MIMO base-stations (BSs) made of arrays of antennas with extreme physical dimensions often with the size of tens of hundreds of wavelengths \cite{xlmimo_mag}.
Such deployments are promising designs to address crowded communication scenarios, integrating the antenna elements with architectural structures of the environment, \textit{e.g.} walls, ceiling, and columns of a stadium, warehouses, or shopping malls \cite{amiri2019message}.
For this reason, the distances between the users and the antenna elements are small compared with the co-located BS design, typically adopted in cellular systems.

Such small distances between the users in conjunction with the very large extent of the antenna array creates spatial non-stationarities on the wireless channel, which drastically changes the signal propagation aspects compared with the conventional massive MIMO scenarios. We investigate two of these aspects.
First, each antenna element experiences different average received power and phase from each user, suggesting the operation under the near-field propagation regime. Hence, the wireless channel is well-modeled by the spherical-wave (SW) model rather than the conventional plane-wave model \cite{lu2021}.
Second, the closeness between the users and the antenna elements results in predominantly line-of-sight (LoS) situation. However, due to the relief and presence of scatterers and obstacles in the environment, it is not accurate to assume that all the radio links experience LoS transmission. Hence, as is discussed in \cite{liu2021}, it is reasonable to assume that part of the radio links are under the LoS regime, while the remaining links are under the non-line-of-sight (NLoS) one.

Under the near-field propagation regime considering the SW model, the array gain is limited \cite{lu2021}. This suggests that asymptotic favorable propagation, which results in the orthogonality between the channel vectors of different users, may also be compromised by the near-field propagation condition too. However, further investigation is needed to support this claim.
The looseness of the favorable propagation with the SW model increases the necessity of scheduling spatially compatible users in order to achieve reasonable downlink performance, while optimizing the expenditure of the scarce radio resources, \textit{e.g.}, transmit power and resource blocks.

In wireless systems with limited resources, resource allocation (RA) techniques are essential to assure minimum quality-of-service (QoS) levels for a higher number of served users.
RA techniques for multi-user MIMO systems are surveyed in \cite{castaneda2017}.
Despite that, further investigation is needed to assess the effectiveness of the conventional resource allocation strategies in crowded multi-user MIMO systems operating under spatially non-stationary wireless channels, which can be attained with XL antenna arrays and high-user density.
The problem of user-scheduling for the XL-MIMO systems with the SW model is addressed in \cite{gonzalez-coma2021}.
The authors propose a scheduling strategy based on the equivalent distance, a measure which combines the distance from the users to the BS and the interference level produced by the other scheduled users.
In \cite{marinello2020,souza2021} the authors propose methods for antenna selection aiming to maximize the energy efficiency and the spectral efficiency, respectively. The proposed methods use different approaches to compute the set of active antennas, including metaheuristic optimization, greedy strategies, and heuristic methods deploying approximate expressions for the performance metrics. However, proposing solutions specially designed for wireless channel models addressing different propagation conditions experienced by the users, while optimizing the expenditure of the scarce radio resources, is still a need.
{In \cite{JCM2022}, authors take advantage of the non-overlapping visibility regions (VR) concept in a high-density, crowded XL-MIMO system to propose a joint random access and user-scheduling protocol.  Such a protocol explores the different VRs of the UEs to improve the access performance, besides seeking UEs with non-overlapping VRs to be scheduled in the same payload data pilot resource.}

\noindent\textit{Contributions}. The contribution of this work is fourfold.
\begin{enumerate}[label={\textit{\roman*})}]

\item We propose a QoS-aware joint user scheduling and power allocation technique for the DL channel of crowded XL-MIMO systems. The proposed technique efficiently solves the formulated RA problem ($\mathcal{P}_0$) with two sequential procedures: the \textit{clique search-based scheduling} (CBS) algorithm for user scheduling ($\mathcal{P}_1$), and optimal power allocation with transmit power budget and minimum achievable rate per user constraints ($\mathcal{P}_2$). Specifically, the CBS algorithm implements a \textit{clique search} procedure on a graph that represents the users into the communication cell and their respective channels. Besides, CBS prioritizes serving the maximum number of users with minimum QoS rather than the common approach of maximizing the DL capacity as in \cite{gonzalez-coma2021} and \cite{taesang2005}.
The adopted optimization goal is specially suitable for crowded scenarios, in which huge numbers of users need to be served by a network structure with limited transmit power budget and resource blocks.

\item Considering the transmit power budget and the minimum required rate per user, we develop an efficient method to check the infeasibility of the power allocation optimization problem ($\mathcal{P}_2$) without needing to calculate an inverse matrix. This method motivates the development of the graph representation used in CBS and reduces significantly the number of operations required to test if the original RA problem is feasible with a given set of scheduled users.

\item Aiming to capture the complexity of the propagation environment with XL antenna arrays, we propose a non-stationary multi-state channel model based on the SW propagation considering that users under LoS and NLoS transmission coexist in the same communication cell. In the proposed model is assumed that users under the LoS and NLoS states experience different propagation aspects both in the multi-path fading model and the path loss rule, extending the model proposed in \cite{ding2016}. Moreover, differently from the model developed in \cite{liu2021}, the proposed SW-based non-stationary multi-state LoS/NLoS model considers the channel state at array level rather than antenna.

\item We extensively evaluate the performance of the proposed CBS in the crowded XL-MIMO scenario under different channel conditions, comparing it to state-of-the-art techniques in terms of achievable sum-rate and number of scheduled users. Moreover, differently from \cite{gonzalez-coma2021} and \cite{taesang2005}, we analyze the individual performance of the scheduled users, characterizing the distribution of the served users along the cell area and according to their channel states.
\end{enumerate}

\vspace{2mm}

\noindent\textit{Notations}. Boldface small $\mathbf{a}$ and capital $\mathbf{A}$ letters represent vectors and matrices, respectively. Capital calligraphic letters $\mathcal{A}$ represent finite sets. $\mathbf{I}_n$ denotes the identity matrix of size $n$.
$\mathbf{0}_n$ denotes the zero column vector of length $n$.
$\{\cdot\}^T$ and $\{\cdot\}^H$ denote, respectively, the transpose and the conjugate transpose operators. $\wp(\cdot)$ denotes the power set operator.

\section{System Model}\label{sec:system-model}

\begin{figure}[b]
\centering
\includegraphics[width=.4\textwidth]{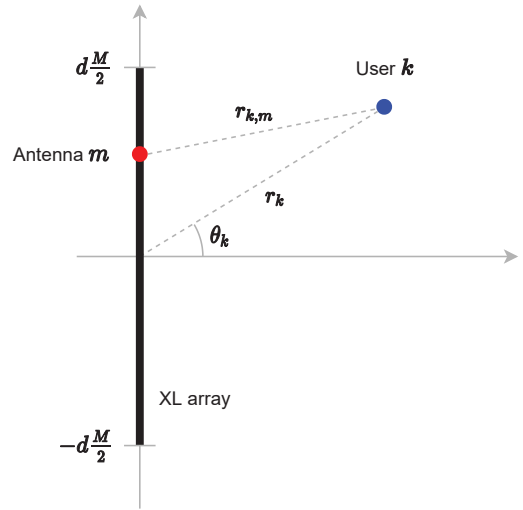}
\vspace{-4mm}
\caption{Diagram of one user at the {XL-MIMO}  communication scenario.}
	\label{fig:communication-scenario}
\end{figure}

In the following, we describe the model of the communication system analyzed in this work.
We consider the DL transmission of a narrowband XL-MIMO system with $K$ users operating in the time-division duplexing (TDD) mode. The BS is equipped with $M$ antennas organized as a uniform linear array with elements spaced by the distance $d$. Therefore, the array has the aperture $D = (M-1)d$. A geometric sketch for a single-user XL-MIMO communication scenario is depicted in Fig. \ref{fig:communication-scenario}.

\subsection{Channel Model}

In this subsection, we formulate the channel model based on the SW model considering that users under LoS and NLoS channel states coexist in the same communication cell. For this reason, we define two channel vectors, one for the LoS channel model, and other for the NLoS one. Then, we define a unified model capturing the multi-state aspect of the proposed channel model.

The LoS channel follows the SW model. Hence, the channel response between the antenna $m$ and the user $k$ is equal to
\begin{equation}
\label{eq:los-response}
a_m (r_k, \theta_k) = \sqrt{\frac{\beta_0^\textsc{LoS}}{r_{k,m}^{\gamma^\textsc{LoS}}}} \exp \left( -j \frac{2\pi}{\lambda} r_{k,m} \right),
\end{equation}
where, $r_k$ is the distance between the user $k$ and the array boresight, $\theta_k$ is the angle formed by the line connecting the user $k$ and the array boresight, $r_{k,m}$ is the distance between the user $k$ and the antenna $m$ (see Fig. \ref{fig:communication-scenario}), $\beta_0^\textsc{LoS}$ is the path-loss attenuation at a reference distance, $\gamma^\textsc{LoS}$ is the path-loss exponent, and $\lambda$ is the carrier wavelength.
Considering \eqref{eq:los-response}, the channel vector $\mathbf{a}_k^\textsc{LoS} \in \complexset^M$ for the LoS channel is equal to
\begin{equation}
\label{eq:los-vector}
\mathbf{a}_k^\textsc{LoS} = \begin{bmatrix} a_1(r_k, \theta_k) & \cdots & a_M(r_k, \theta_k) \end{bmatrix}^T
\end{equation}

Differently, the NLoS channel follows the i.i.d. Rayleigh-fading model with the path-loss computed independently for each antenna due to the variation of the average received power across the large-aperture XL-MIMO array \cite{marinello2020}. The path-loss of the NLoS radio link between user $k$ and the antenna $m$ is equal to
\begin{equation}
\beta_m(r_k, \theta_k) = \frac{\beta_0^\textsc{NLoS}}{r_{k,m}^{\gamma^\textsc{NLoS}}}
\end{equation}
where $\beta_0^\textsc{NLoS}$ is the path-loss attenuation at a reference distance and $\gamma^\textsc{NLoS}$ is the path-loss exponent.
Hence, the channel vector $\mathbf{a}_k^\textsc{NLoS} \in \complexset^M$ for the NLoS channel is defined such that
\begin{equation}
\label{eq:nlos-channel} \mathbf{a}_k^\textsc{NLoS} \cscgdist{\mathbf{0}_M}{\boldsymbol{\Sigma}}
\end{equation}
with the diagonal covariance matrix $\boldsymbol{\Sigma}$ containing the path-loss coefficients w.r.t. all the antenna elements, \textit{i.e.}
\begin{equation}
\label{eq:covmat}
\boldsymbol{\Sigma} = \diag{\begin{bmatrix} \beta_1(r_k, \theta_k) & \cdots & \beta_M(r_k, \theta_k) \end{bmatrix}^T}\\
\end{equation}
Note that the uncorrelated channel assumption in \eqref{eq:covmat} is justified since the antenna elements are separated by a distance $d \geq \lambda/2$.

\vspace{2mm}

\noindent\textit{Definition 1.} Let $x_k \in \{0,1\}$ be the \textit{channel state indicator} associated to the user $k$, equal to 1 if the channel is under the LoS state, or 0 if it is under the NLoS state. To capture the influence of topographic features associated to the communication cell on the channel state, \textit{e.g.}, the effect of relief, as well as the spatial configuration of scatterers and obstacles, the indicator is modelled as a random variable. Therefore, $x_k \sim f_{x_k \mid r_k, \theta_k}(x \mid r_k, \theta_k)$, where $f_{x_k \mid r_k, \theta_k} : \{0,1\} \rightarrow \realset_+^*$ is the conditional probability mass function (pmf) that depends on the position of the user $k$ into the cell.

With the definition of the channel state indicator, as well as the LoS and NLoS channel vectors, we can define the \textit{multi-state channel vector} $\mathbf{a}_k \in \complexset^M$ as
\begin{equation}
\label{eq:mscv}
\mathbf{a}_k = x_k \mathbf{a}_k^\textsc{LoS} + (1 - x_k) \mathbf{a}_k^\textsc{NLoS}
\end{equation}
Notice that, when $x_k = 1$, eq. \eqref{eq:mscv} is equal to the LoS channel vector. On the other hand, when $x_k = 0$, the channel vector of user $k$ is equal to the NLoS channel vector. Hence, users with different channel states may coexist in the same communication cell, depending on the definition of the state indicator pmf, $f_{x_k \mid r_k, \theta_k}$.

In this sense, for the sake of simplicity and to enable evaluating the proposed techniques in a variety of channel scenarios, in the remainder of this work we consider that the channel state indicators follow a Bernoulli random distribution with parameter $\rho$, namely the \textit{LoS probability}. Therefore, the conditional pmf results:
\begin{equation}
f_{x_k \mid r_k, \theta_k}(x \mid r_k, \theta_k) = \rho^x (1 - \rho)^{1 - x}
\end{equation}
where $x \in \{0,1\}$ and $\forall k \in \{1,\dots,K\}$.

\subsection{Signal Model}

Now, we define the model for the signal received by the users. Let $\mathcal{K} \subseteq \{1,\dots,K\}$ be the set of scheduled users. The transmitted signal by the BS is equal to
\begin{equation}
\textbf{z} = \sum_{k \in \mathcal{K}} \sqrt{p_k} s_k \mathbf{f}_k
\end{equation}
where $p_k \geq 0$ is the power allocated for the user $k$, $s_k$ such that $\expval{\abs{s_k}^2} = 1$ is the signal intended for the user $k$, and $\mathbf{f}_k$ such that $\norm{\mathbf{f}_k}{2}^2 = 1$ is the precoding vector computed for the user $k$.
The received signal by the user $k \in \mathcal{K}$ is equal to
\begin{equation}
\label{eq:received-signal}
y_k = \sqrt{p_k} s_k \mathbf{a}_k^H \mathbf{f}_k + \sum_{i \in \mathcal{K}\backslash k} \sqrt{p_i} s_i \mathbf{a}_k^H \mathbf{f}_i + w_k
\end{equation}
where $w_k \cscgdist{0}{\sigma_w^2}$ is the additive white Gaussian noise sample.
Given the received signal in \eqref{eq:received-signal}, the signal-to-interference-plus-noise ratio (SINR) calculated for the user $k \in \mathcal{K}$ is equal to \cite{gonzalez-coma2021}
\begin{equation}
\label{eq:sinr}
\mathrm{SINR}_k = \frac{p_k \abs{\mathbf{a}_k^H \mathbf{f}_k}^2}{\sum_{i \in \mathcal{K}\backslash k} p_i \abs{\mathbf{a}_k^H \mathbf{f}_i}^2 + \sigma_w^2}
\end{equation}

Without loss of generality, let $k = 1,\dots,\abs{\mathcal{K}}$ be the indices of the scheduled users. Then, the channel matrix with the channel vectors of all the scheduled users is defined as $\mathbf{A} \in \complexset^{M \times |\mathcal{K}|}$ such that $\mathbf{A} = \begin{bmatrix} \mathbf{a}_1 & \cdots & \mathbf{a}_{\abs{\mathcal{K}}} \end{bmatrix}$. For the sake of simplicity, we consider perfect channel state information available at the transmitter (CSIT), and that $\abs{\mathcal{K}} \leq M$ and $\mathrm{rank} \; \mathbf{A} = \abs{\mathcal{K}}$, the BS transmits the DL signal using the zero-forcing (ZF) precoder. Hence, the precoding vector for each user $k \in \mathcal{K}$ is equal to
\begin{equation}
\label{eq:precoding-vector}
\mathbf{f}_k^\textsc{zf} = \frac{\mathbf{A} \left( \mathbf{A}^H \mathbf{A} \right)^{-1} \mathbf{e}_k}{\left[ \left( \mathbf{A}^H \mathbf{A} \right)^{-1} \right]_{k,k}^{\frac{1}{2}}}
\end{equation}
where
$\mathbf{e}_k \in \{0,1\}^{\abs{\mathcal{K}}}$ is the $k$-th vector of the standard basis of the $\abs{\mathcal{K}}$-dimensional Euclidean space.

Considering that the ZF precoder mitigates the inter-user interference (IUI), \textit{i.e.}, $\mathbf{a}_i^H \mathbf{f}_j = 0, \forall i,j \in \mathcal{K}, i \neq j$, substituting \eqref{eq:precoding-vector} in eq. \eqref{eq:sinr} results in the SINR calculated for user $k$ using the ZF precoder,
\begin{equation}
\label{eq:sinr-zf}
\mathrm{SINR}_k^\textsc{zf} = \frac{p_k}{\sigma_w^2 \left[ \left( \mathbf{A}^H \mathbf{A} \right)^{-1} \right]_{k,k}}
\end{equation}
Using the Shannon's equation, the achievable rate of the user $k \in \mathcal{K}$ using the ZF precoder is equal to
\begin{align}
R_k^\textsc{zf} & = \log_2 \left( 1 + \mathrm{SINR}_k^\textsc{zf} \right) \notag\\
& = \log_2 \left( 1 + \frac{p_k}{\sigma_w^2 \left[ \left( \mathbf{A}^H \mathbf{A} \right)^{-1} \right]_{k,k}} \right) \label{eq:achievable-rate}
\end{align}
Observing \eqref{eq:achievable-rate}, we note that the achievable rate of the user $k$ not only depends on its respective allocated power and the noise power, but also on the overall set of scheduled users and their respective channel vectors. For this reason, the user scheduling process is crucial to attain reasonable performance levels.

Let $\overline{R}_k \in \realset_+, \forall k \in \mathcal{K}$ be the minimum achievable rate that the BS must serve to user $k$. Given the set of scheduled users $\mathcal{K}$, the maximum downlink achievable sum-rate of the XL-MIMO system using the ZF precoder is succeed with the allocated powers that solve the following optimization problem:
\begin{subequations}
\label{eq:maximum-achievable-sum-rate}
\begin{align}
\{p_k^*\}_{k \in \mathcal{K}} = \; & \underset{\{p_k\}_{k \in \mathcal{K}}}{\argmax} \quad \sum_{k \in \mathcal{K}} R_k^\textsc{zf}\\
& \mathrm{subject\;to} \quad R_k^\textsc{zf} \geq \overline{R}_k, \; \forall k \in \mathcal{K}\\
& \qquad\qquad\quad\, \sum_{k \in \mathcal{K}} p_k \leq P_{\max}\\
& \qquad\qquad\quad\;\, p_k \geq 0, \forall k \in \mathcal{K}
\end{align}
\end{subequations}
where $P_{\max}$ is the maximum power available for DL transmission, and the achievable rates $R_k^\textsc{zf}, \forall k \in \mathcal{K}$ are given by eq. \eqref{eq:achievable-rate}.
Since eq. \eqref{eq:maximum-achievable-sum-rate} is equivalent to the optimization problem of allocating power on independent parallel Gaussian channels, the set of powers $\{p_k^*\}_{k \in \mathcal{K}}$ that solve it follows the water-filling distribution \cite{palomar2005}.

\section{User Scheduling: Problem Formulation}\label{sec:user-scheduling-problem-formulation}

In this section, we introduce the formulation of the studied user scheduling problem. The optimization problem of joint DL user scheduling and power allocation with individual minimum achievable rate constraints and transmit power budget can be defined as
\begin{subequations}
\label{eq:original-problem}
\begin{align}
\mathcal{P}_0: \quad \underset{\mathcal{K}, \{p_k\}_{k \in \mathcal{K}}}{\mathrm{maximize}} \quad & \sum_{k \in \mathcal{K}} R_k^\textsc{zf}\\
\label{eq:original-problem-minimum-rates}
\mathrm{subject\;to} \quad & R_k^\textsc{zf} \geq \overline{R}_k, \; \forall k \in \mathcal{K}\\
\label{eq:original-problem-maximum-power}
& \sum_{k \in \mathcal{K}} p_k \leq P_{\max}\\
\label{eq:original-problem-set-scheduled-users-domain}
& \mathcal{K} \subseteq \{ 1,\dots,K \}\\
\label{eq:original-problem-powers-domain}
& p_k \geq 0, \forall k \in \mathcal{K}
\end{align}
\end{subequations}
The constraints \eqref{eq:original-problem-minimum-rates} ensure that all the scheduled users are served with a minimum achievable rate. Moreover, the constraint \eqref{eq:original-problem-maximum-power} ensures that the DL transmitted power does not exceed $P_{\max}$. Finally, the constraints \eqref{eq:original-problem-powers-domain} and \eqref{eq:original-problem-set-scheduled-users-domain} define the domain of the optimization variables.

The optimization problem $\mathcal{P}_0$ is concave in the variables $\{p_k\}_{k \in \mathcal{K}}$, but not in the variable $\mathcal{K}$. For this reason, it isn't possible to solve $\mathcal{P}_0$ optimally with standard convex optimization tools. An alternative path to reach a sub-optimal solution is to split $\mathcal{P}_0$ into two sub-problems in which each variable is optimized independently. We discuss this strategy in the sequel.

Let $g: \wp(\{ \mathbf{a}_k \}_{k=1}^K) \rightarrow \realset_+$ be a function that measures the \textbf{spatial compatibility} between users from their channel vectors. The {spatial compatibility} quantifies how efficiently these channel vectors can be separated in space. Examples of spatial compatibility metrics are the \textit{condition number} and the \textit{null-space projection} of the channel matrix. Since there exists a correspondence between spatial compatible users and the precoding performance, optimizing a spatial compatibility metric is a promising path to obtain a good set of scheduled users \cite{castaneda2017}.
The generic user scheduling problem solved by maximizing a spatial compatibility metric is called \textit{user grouping}, and can be formulated as
\begin{subequations}
\label{eq:sub-problem-set-scheduled-users}
\begin{align}
\mathcal{P}_1: \quad \mathcal{K}^* = \; & \underset{\mathcal{K}}{\argmax} \quad g(\{\mathbf{a}_k\}_{k \in \mathcal{K}})\\
& \mathrm{subject\;to} \quad \mathcal{K} \subseteq \{ 1,\dots,K \}
\end{align}
\end{subequations}
The optimization problem $\mathcal{P}_1$ is an NP-complete combinatorial problem solved only by exhaustive search. Since, in crowded XL-MIMO systems, the number of users into the communication cell is high, the size of the solution space of $\mathcal{P}_1$ scales quickly. Hence, in such a case it is impractical to solve the user grouping problem in feasible time. For this reason, Section \ref{sec:user-scheduling-clique-search} develops an effective, quasi-optimal and computationally efficient method to carry out the user scheduling in crowded XL-MIMO scenarios.

Given the set of scheduled users $\mathcal{K}^*$, the optimal set of allocated powers $\{p_k^*\}_{k \in \mathcal{K}^*}$ can be calculated by solving the following optimization
sub-problem:
\begin{subequations}
\label{eq:sub-problem-powers}
\begin{align}
\mathcal{P}_2: \quad \{p_k^*\}_{k \in \mathcal{K}^*} = \; & \underset{\{p_k\}_{k \in \mathcal{K}^*}}{\argmax} \quad \sum_{k \in \mathcal{K}^*} R_k^\textsc{zf}\\
& \mathrm{subject\;to} \;\; R_k^\textsc{zf} \geq \overline{R}_k, \; \forall k \in \mathcal{K}^*\\
&\label{eq:sub-problem-powers-maximum-power}
\qquad\qquad\;\;\, \sum_{k \in \mathcal{K}^*} p_k \leq P_{\max}\\
& \qquad\qquad\;\;\;\, p_k \geq 0, \forall k \in \mathcal{K}^*
\end{align}
\end{subequations}
The optimization problem $\mathcal{P}_2$ is identical to \eqref{eq:maximum-achievable-sum-rate} and, if feasible, it can be solved optimally by the water-filling solution \cite{palomar2005}. A simple way to check the feasibility of $\mathcal{P}_2$ is presented in the following.

\vspace{2mm}

\noindent\textit{Remark 1.} The optimization problem $\mathcal{P}_2$ is feasible if and only if the sum of the minimum allocated powers necessary to serve each user with its respective minimum achievable rate do not exceed $P_{\max}$, \textit{i.e.},
\begin{equation}
\label{eq:power-allocation-feasibility}
\sum_{k \in \mathcal{K}^*} \overline{p}_k = \sigma_w^2 \sum_{k \in \mathcal{K}^*} \left( 2^{\overline{R}_k} - 1 \right) \left[ \left( \mathbf{A}^H \mathbf{A} \right)^{-1} \right]_{k,k}
\leq P_{\max}
\end{equation}
where $\overline{p}_k$ is the minimum power required to serve user $k$ with the achievable rate $\overline{R}_k$, obtained from eq. \eqref{eq:achievable-rate}.

Considering that the feasibility criterion in eq. \eqref{eq:power-allocation-feasibility} is satisfied, the solution of $\mathcal{P}_2$ is given by
\begin{equation}
p_k^* = \max \left(\overline{p}_k, \;  \mu - \sigma_w^2 \left[ \left( \mathbf{A}^H \mathbf{A} \right)^{-1} \right]_{k,k} \right)
\end{equation}
$\forall k \in \mathcal{K}$, where $\mu$ is a constant called \textit{water-level}. Moreover, in order to meet the constraint \eqref{eq:sub-problem-powers-maximum-power} with equality, the optimal water-level can be obtained satisfying:
\begin{equation}
\sum_{k \in \mathcal{K}^*} \max \left(\overline{p}_k, \;  \mu - \sigma_w^2 \left[ \left( \mathbf{A}^H \mathbf{A} \right)^{-1} \right]_{k,k} \right) - P_{\max} = 0
\end{equation}
which can be easily solved by a root-finding algorithm \cite{palomar2005}.

A suboptimal solution of the original problem $\mathcal{P}_0$ can be obtained by sequentially solving $\mathcal{P}_1$ and $\mathcal{P}_2$. Since the two optimization variables are decoupled in the formulated sub-problems, the set of scheduled users $\mathcal{K}^*$ may result in an infeasible power allocation policy. In such case, the set $\mathcal{K}^*$ must be altered in order to enable the power allocation.

\begin{figure}[b]
\centering
\includegraphics[width=\columnwidth]{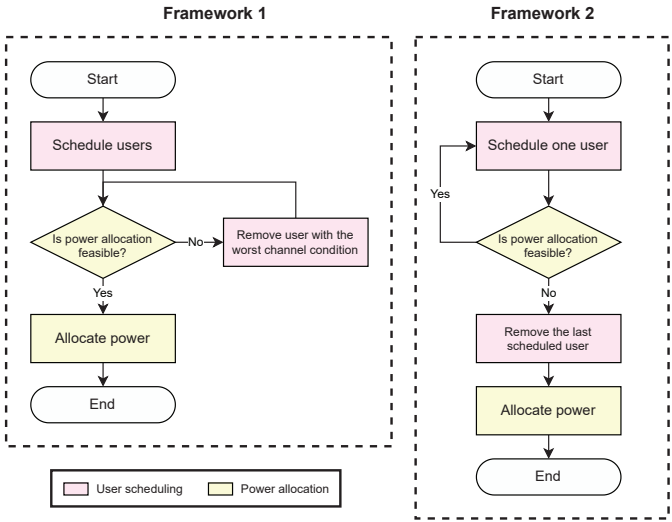}
\caption{Flowcharts of the two distinct frameworks adopted for joint user scheduling and power allocation procedure by sequentially solving $\mathcal{P}_1$ and $\mathcal{P}_2$. Note that, in Framework 2, the feasibility of the power allocation problem is checked at every iteration of the user scheduling procedure. On the other hand, in Framework 1, the power allocation feasibility check is carried out more than once only if the set of scheduled users results in an infeasible power allocation problem.}
\label{fig:frameworks-flowcharts}
\end{figure}

To solve the user scheduling problem by sequentially solving the sub-problems $\mathcal{P}_1$ and $\mathcal{P}_2$, we propose the two distinct frameworks presented in Fig. \ref{fig:frameworks-flowcharts}.
In the \textit{Framework 1}, firstly the set of scheduled users is computed. Next, if the power allocation with the calculated set of users is infeasible, the users with the worst channel condition are removed until feasibility is reached. Finally, the power allocation is carried out.
Differently, in the \textit{Framework 2}, users are scheduled iteratively until the power allocation problem becomes infeasible. When an infeasible set of users is reached, the last scheduled user is removed, then power allocation procedure is carried out. From an implementation perspective, in general, solutions that fit into Framework 2 demand higher computational complexity than those fitting into Framework 1. This is due to the power allocation feasibility test carried out at every iteration, which commonly requires the calculation of the precoding vectors.
Although in Framework 1 this feasibility test is carried out as well, if the user scheduling procedure is carefully designed, the number of tests can be drastically reduced, increasing its computational advantage w.r.t. Framework 2.

\subsection{\texorpdfstring{$\mathcal{P}_2$}{P2} Infeasibility Test}

In this subsection, we present an efficient method to check the infeasibility of the optimization problem \eqref{eq:sub-problem-powers} without needing to calculate the inverse matrix $\left( \mathbf{A}^H \mathbf{A} \right)^{-1}$. This method motivates the development of the graph representation used in the proposed scheduling algorithm and reduces significantly the number of operations required to test the feasibility of the set of scheduled users.

Let $R_k^{\textsc{su}}$ be the single user capacity of the user $k$ calculated by
\begin{equation}
R_k^{\textsc{su}} = \log_2 \left( 1 + \frac{p_k \norm{\mathbf{a}_k}{2}^2}{\sigma_w^2} \right)
\end{equation}
The single user capacity is the achievable rate if, during the DL, the BS transmits only the signal of user $k$.
Following this definition, the minimum required power $\overline{p}_k^{\textsc{su}}$ for user $k$ to experience its minimum achievable rate $\overline{R}_k$ is equal to
\begin{equation}
\overline{p}_k^{\textsc{su}} = \frac{\sigma_w^2}{\norm{\mathbf{a}_k}{2}^2} \left( 2^{\overline{R}_k} - 1 \right)
\end{equation}

\vspace{2mm}
\noindent\textit{Lemma 1.} For any set of scheduled users $\mathcal{K} \subseteq \{ 1,\dots,K \}$, if the sum of the minimum powers required to equals the single user capacity of each scheduled user to its minimum achievable rate is equal to or greater than $P_{\max}$, \textit{i.e.}, $\sum_{k \in \mathcal{K}} \overline{p}_k^{\textsc{su}} \geq P_{\max}$, the optimization problem \eqref{eq:sub-problem-powers} is infeasible.
Such condition is sufficient but not necessary to confirm the infeasbility of $\mathcal{P}_2$. If $\sum_{k \in \mathcal{K}} \overline{p}_k^{\textsc{su}} < P_{\max}$, the feasibility or infeasibility of $\mathcal{P}_2$ can only be proved by checking whether eq. \eqref{eq:power-allocation-feasibility} holds.

\vspace{2mm}

\noindent\textit{Proof.} The effective channel gain obtained by user $k$ with the ZF precoder is upper-bounded by \cite{lu2022}
\begin{equation}
\label{eq:upper-bound-zf-channel-gain}\small
\norm{\mathbf{a}_k}{2}^2 \geq \left[ \left( \mathbf{A}^H \mathbf{A} \right)^{-1} \right]_{k,k}^{-1} = \| \mathbf{a}_k \|_2^2 -  \mathbf{a}_k^H \Breve{\mathbf{A}}_k \left( \Breve{\mathbf{A}}_k^H \Breve{\mathbf{A}}_k \right)^{-1} \Breve{\mathbf{A}}_k^H \mathbf{a}_k
\end{equation}
where $\Breve{\mathbf{A}}_k \in \complexset^{M \times \abs{\mathcal{K}}-1}$ is the channel matrix with the channel vectors of all the scheduled users, except for $k$, \textit{i.e.},
\begin{equation}
\Breve{\mathbf{A}}_k = \begin{bmatrix} \mathbf{a}_1 & \cdots & \mathbf{a}_{k-1} & \mathbf{a}_{k+1} & \cdots & \mathbf{a}_{\abs{\mathcal{K}}} \end{bmatrix}
\end{equation}
Since $\mathbf{A}$ has full rank, $\left( \Breve{\mathbf{A}}_k^H \Breve{\mathbf{A}}_k \right)^{-1}$ is positive definite and, consequently, the equality in eq. \eqref{eq:upper-bound-zf-channel-gain} is obtained if and only if $\Breve{\mathbf{A}}_k^H \mathbf{a}_k = \mathbf{0}_{\abs{\mathcal{K}}-1}$, \textit{i.e.}, the channel vector of user $k$ is orthogonal to the channel vectors of all the other users. Therefore, we obtain the following relationship between the sum of the minimum allocated powers required to attain the minimum achievable rates of the scheduled users,
\begin{equation}
\sum_{k \in \mathcal{K}} \overline{p}_k \geq \sum_{k \in \mathcal{K}} \overline{p}_k^{\textsc{su}}
\end{equation}
Accordingly, assuming that it is impossible to get perfectly orthogonal channel vectors, if $\sum_{k \in \mathcal{K}} \overline{p}_k^{\textsc{su}} \geq P_{\max}$ we have that $\sum_{k \in \mathcal{K}} \overline{p}_k > P_{\max}$, indicating that $\mathcal{K}$ is an infeasible set of scheduled users for the optimization problem $\mathcal{P}_2$. \hfill $\blacksquare$

\section{User Scheduling Based on Clique Search}\label{sec:user-scheduling-clique-search}

In this section, first we introduce the concept of undirected vertex-weighted graph (UWG) model for modeling the interference between the users in the proposed user scheduling XL-MIMO operating under a non-stationary multi-state LoS and NLoS channels.
Then, we formulate a clique search problem on the UWG to solve the user scheduling and power allocation sub-problems $\mathcal{P}_1$ and $\mathcal{P}_2$ proposed in Section \ref{sec:user-scheduling-problem-formulation}.

\subsection{Undirected Vertex-Weighted Graph Model}

Let $\mathcal{G} = (\mathcal{V}, \mathcal{E})$ be an UWG, where $\mathcal{V} = \{ v_1,\dots,v_V \}$ is the set with the graph vertices such that $\abs{\mathcal{V}} = V$, and $\mathcal{E} \subseteq \set{\{v_i, v_j\} \mid v_i,v_j \in \mathcal{V}, v_i \neq v_j}$ is the set with the graph edges. Let $\mathbf{E} \in \{0,1\}^{V \times V}$ be the adjacency matrix of the graph $\mathcal{G}$ such that
\begin{equation}
[\mathbf{E}]_{i,j} = \begin{cases}
1, & \text{if} \; \{v_i, v_j\}\in \mathcal{E}\\
0, & \text{otherwise}
\end{cases}
\end{equation}
The vertex weight function $\omega : \mathcal{V} \rightarrow \realset$ characterizes the weight of each vertex $v_i \in \mathcal{V}$.

\begin{figure}[b]
\centering
\includegraphics[width=\linewidth]{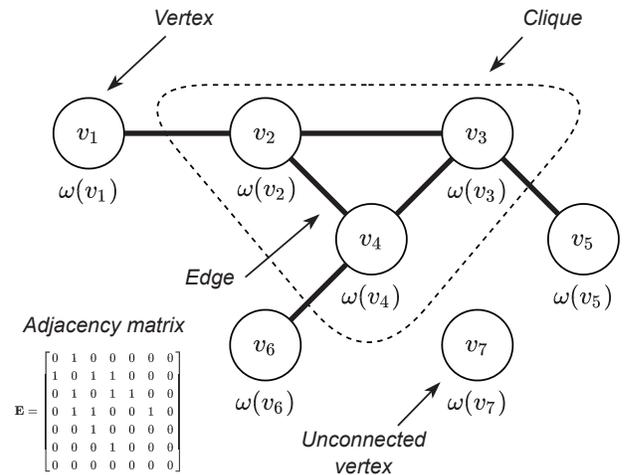}
\caption{UWG representation of a hypothetical XL-MIMO system with $K = 7$ users and its equivalent adjacency matrix. The vertices $v_2$, $v_3$, and $v_4$ form a clique in this graph.}
\label{fig:graph-representation}
\end{figure}

{In our work, the UWG $\mathcal{G}$ represents the users into the communication cell and the orthogonality relationship between their respective channel vectors. Each user $k \in \{ 1,\dots,K \}$ is represented by a vertex $v_k$. Moreover, the edges $\mathcal{E}$ are described by the adjacency matrix constructed from the channel vectors according to the $\epsilon$-orthogonal rule,
\begin{equation}
[\mathbf{E}]_{i,j} = \begin{cases}
0, & \text{if} \; i = j\\
I\left( \frac{|\mathbf{a}_i^H \mathbf{a}_j|}{\norm{\mathbf{a}_i}{2} \norm{\mathbf{a}_j}{2}} < \epsilon \right), & \text{otherwise}
\end{cases}
\end{equation}
$\forall i,j \in \{ 1,\dots,K \}$, where $I(\cdot)$ is the indicator function, and $\epsilon$ represents the \textit{admissibility for channel orthogonality}. This method of defining the graph edges makes that only vertices that represent users with quasi-orthogonal channel vectors to be connected.
Finally, the weight of vertex $k$ is defined as the minimum power required for user $k$ to achieve a single user capacity equal to its minimum achievable rate, \textit{i.e.},
\begin{equation}
\omega(v_k) = \overline{p}_k^{\textsc{su}} = \frac{\sigma_w^2}{\norm{a_k}{2}^2} \left( 2^{\overline{R}_k} - 1 \right), \; \forall k \in \{ 1,\dots,K \}
\end{equation}
Fig. \ref{fig:graph-representation} depicts a diagram of an UWG representation of a hypothetical XL-MIMO system with $K = 7$ users and its equivalent adjacency matrix.}

\textit{Definition 2.} The subgraph $\mathcal{G}' = (\mathcal{V}', \mathcal{E}')$ is called a \textit{clique} if the vertices in $\mathcal{V}'$ are mutually adjacent to one another, {\it \textit{i.e.}}, $\mathcal{E}' = \set{\{v_i, v_j\} \mid v_i, v_j \in \mathcal{V}', v_i \neq v_j}$. Moreover, the number of vertices in the clique is called \textit{clique number} \cite{gross2014}.

\subsection{User Scheduling Based on Clique Search}

Now, we formulate the user scheduling process as a clique search problem in the UWG $\mathcal{G}$ that represents the XL-MIMO system. Let $\mathcal{C}_\mathcal{G}$ be the set of all the cliques of the graph $\mathcal{G}$. Our aim is to find the clique with the largest clique number with sum of the weights less than or equal to $P_{\max}$. This clique search problem can be written in the form
\begin{subequations}
\label{eq:problem-clique-search}
\begin{align}
\label{eq:problem-clique-objective}
\mathcal{P}_3: \qquad \mathcal{G}^* = \; & \underset{\mathcal{G}'}{\argmax} \quad \abs{\mathcal{V}'}\\
\label{eq:problem-clique-max-sum-weights}
& \mathrm{subject\;to} \quad \sum_{v_k \in \mathcal{V}'} \omega (v_k) \leq P_{\max}\\
\label{eq:problem-clique-graph-domain}
& \qquad\qquad\quad\;\;\; \mathcal{G}' = (\mathcal{V}', \mathcal{E}') \in \mathcal{C}_{\mathcal{G}}
\end{align}
\end{subequations}
The constraint \eqref{eq:problem-clique-max-sum-weights} ensures, via the sum of the weights of the clique vertices, that the set of scheduled users represented by the vertices is in the limit of $\mathcal{P}_2$ infeasibility. Moreover, the constraint \eqref{eq:problem-clique-graph-domain} ensures that $\mathcal{G}^*$ is a clique of the graph $\mathcal{G}$.
Typically, clique problems are NP-complete, requiring high computational effort to be solved in large graphs. Specifically, the problem $\mathcal{P}_3$ can be solved by clique enumeration \cite{gross2014}, which consists on listing all the possible cliques in the graph $\mathcal{G}$, evaluating which one simultaneously satisfies \eqref{eq:problem-clique-objective} and \eqref{eq:problem-clique-max-sum-weights}. The search space of this procedure can be reduced by evaluating the constraint \eqref{eq:problem-clique-max-sum-weights} during each iteration of the clique enumeration. However, the worst-time complexity of this procedure is still the same as the original one. For this reason, we propose a low-complexity procedure to calculate a near-optimal solution of $\mathcal{P}_3$.

Algorithm \ref{alg:clique-search} presents a method to find a near-optimal solution for the clique search problem $\mathcal{P}_3$. We call this algorithm as \textit{clique search-based scheduling} (CBS). In this pseudocode, the operator $\mathrm{neighbors}: \mathcal{V} \rightarrow \wp(\mathcal{V})$ returns the set of vertices that have edges with the input vertex $v_k$, \textit{i.e.}, $\mathrm{neighbors}(v_k) = \set{v_i \mid \{v_k, v_i\} \in \mathcal{E}}$.

\begin{algorithm}[b]
\KwIn{The set of graph vertices, $\mathcal{V}$}
\KwOut{The set of clique vertices, $\mathcal{V}'$}

$v_k \gets \underset{v_i \in \mathcal{V}}{\argmin} \; \omega(v_i)$\;
$\mathcal{V}'^{(0)} \gets \{v_k\}$\;
$\mathcal{N}^{(0)} \gets \mathrm{neighbors}(v_k)$\;
$\Omega^{(0)} \gets \omega(v_i)$\;
$n \gets 0$\;

\Repeat{$\mathcal{N}^{(n)} = \emptyset$}{

$v_k \gets \underset{v_i \in \mathcal{N}^{(n)}}{\argmin} \; \omega(v_i)$\;
$\mathcal{V}'^{(n+1)} \gets \mathcal{V}'^{(n)} \cup \{v_k\}$\;
$\mathcal{N}^{(n+1)} \gets \mathcal{N}^{(n)} \cap \mathrm{neighbors}(v_k)$\;
$\Omega^{(n+1)} \gets \Omega^{(n)} + \omega(v_i)$\;

\If{$\Omega^{(n+1)} \geq P_{\max}$}{
exit loop\;
}

$n \gets n + 1$\;
}

$\mathcal{V}' \gets \mathcal{V}'^{(n)}$\;
\caption{\small \textbf{CBS} -- Greedy algorithm to solve the clique search problem $\mathcal{P}_3$.}
\label{alg:clique-search}
\end{algorithm}

The greedy algorithm starts by finding the vertex that requires the least weight and adding it to the clique. The neighboring vertices of the first vertex constitute the clique neighborhood, computed in the line 3.
In the loop beginning at line 6, the vertex of the clique neighborhood with the least weight is added to the clique. Next, in line 9, the clique neighborhood is updated with the vertices that are simultaneously neighbors of all the clique vertices. This loop repeats until the sum of the weights in the clique is less than or equal to $P_{\max}$, or if there are no more vertices in the clique neighborhood. The set of scheduled users is calculated from the output of the Algorithm \ref{alg:clique-search} by $\mathcal{K} = \set{k \in \{ 1,\dots,K \} \mid v_k \in \mathcal{V}'}$. It is important to mention that, since the feasibility of the set of scheduled users w.r.t. the power allocation problem $\mathcal{P}_2$ is not checked in the CBS algorithm, we can't guarantee that the users in $\mathcal{K}$ can be scheduled satisfying the power budget and minimum achievable rate constraints simultaneously. For this reason, we need a procedure to verify the feasibility of $\mathcal{K}$ w.r.t. the problem $\mathcal{P}_2$ and remove users from the set if the power allocation is infeasible. We describe the adopted approach in the following.

\subsection{Obtaining a Feasible Set of Scheduled Users}\label{sec:obtaining-feasible-set-scheduled-users}

Aiming to obtain a feasible set of scheduled users from the CBS algorithm, we adopt the \textit{user removal} procedure described in Algorithm \ref{alg:user-removal}.
In this procedure, the user with the lowest channel power is removed from the set of scheduled users until the power allocation problem $\mathcal{P}_2$ becomes feasible. In line 3, the minimum allocated powers necessary to serve the scheduled users with their minimum achievable rates is calculated. Since the output of the CBS algorithm is an infeasible set of scheduled users, the sum of these powers is greater than $P_{\max}$. In order to obtain a feasible set of scheduled users, in lines 5-8, the user with the lowest channel power is removed; then the minimum powers are recalculated with the new reduced set of users. This procedure repeats until the power allocation feasibility is confirmed according to the condition described in Remark 1, and evaluated in line 9.

\begin{algorithm}[b]
\KwIn{The set of scheduled users, $\mathcal{K}$, the minimum achievable rates, $\set{\overline{R}_k}_{k \in \mathcal{K}}$, and the channel vectors, $\set{\mathbf{a}_k}_{k \in \mathcal{K}}$}
\KwOut{The feasible set of scheduled users, $\mathcal{K}'$}
    
$\mathcal{K}' \gets \mathcal{K}$\;

$\mathbf{A} \gets \begin{bmatrix} \mathbf{a}_1 & \cdots & \mathbf{a}_{|\mathcal{K'}|}  \end{bmatrix}$\;

$\overline{p}_k = \sigma_w^2 \left( 2^{\overline{R}_k} - 1 \right) \left[ \left( \mathbf{A}^H \mathbf{A} \right)^{-1} \right]_{k,k}, \forall k \in \mathcal{K}'$\;
    
\Repeat{$\sum_{k \in \mathcal{K'}} \overline{p}_k \leq P_{\max}$}{

$k^* \gets \underset{k}{\argmin} \; \norm{\mathbf{a}_k}{2}^2, \; k \in \mathcal{K}'$\;

$\mathcal{K}' \gets \mathcal{K}' \backslash k^*$;

$\mathbf{A} \gets \begin{bmatrix} \mathbf{a}_1 & \cdots & \mathbf{a}_{k^*-1} & \mathbf{a}_{k^*+1} & \cdots & \mathbf{a}_{|\mathcal{K'}|}  \end{bmatrix}$\;

$\overline{p}_k = \sigma_w^2 \left( 2^{\overline{R}_k} - 1 \right) \left[ \left( \mathbf{A}^H \mathbf{A} \right)^{-1} \right]_{k,k}, \forall k \in \mathcal{K}'$\;

}

\caption{\small \textbf{User removal}: obtaining a feasible set of scheduled users}
\label{alg:user-removal}
\end{algorithm}

\begin{figure}[t]
\centering
\includegraphics[width=.75\columnwidth]{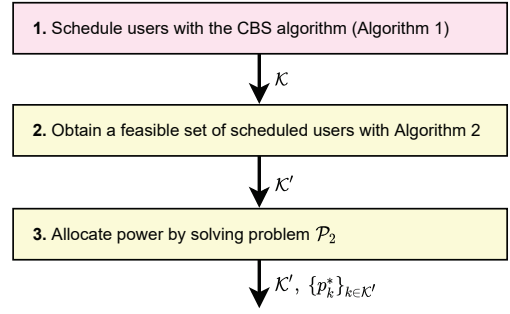}
\vspace{-2mm}
\caption{Flowchart for the proposed joint user scheduling and power allocation technique constituted by the CBS algorithm (user scheduling), a user removal algorithm (set of feasible users), and optimal power allocation with minimum achievable rate constraints.}
\label{fig:solution-flowchart}
\end{figure}

Finally, with a feasible set of scheduled users, we can carry out power allocation satisfying both the \textit{transmit power budget} and the \textit{minimum achievable rate} constraints.
Fig. \ref{fig:solution-flowchart} sketches out the whole proposed technique for joint user scheduling and power allocation in crowded XL-MIMO systems.

\subsection{Scheduling Users Analyzing Their Channel Powers}

Now, we describe a simple but effective approach to schedule users based on the powers of the channel vectors.
Let $P_k$ be the power of the channel vector of user $k$, calculated from the multi-state channel vector in eq. \eqref{eq:mscv} by
\begin{equation}
P_k = \norm{\mathbf{a}_k}{2}^2
\end{equation}
The power of the channel vector is a suitable measure of the channel quality for a given user, resulting from its distance w.r.t. the BS array and its channel state. Moreover, by inspecting eqs. \eqref{eq:power-allocation-feasibility} and \eqref{eq:upper-bound-zf-channel-gain}, one can see that the minimum allocated power necessary to serve user $k \in \mathcal{K}$ with its respective minimum achievable rate is inversely proportional to $P_k$,
\begin{equation}
\overline{p}_k = \frac{\sigma_w^2 \left( 2^{\overline{R}_k} - 1 \right)}{P_k -  \mathbf{a}_k^H \Breve{\mathbf{A}}_k \left( \Breve{\mathbf{A}}_k^H \Breve{\mathbf{A}}_k \right)^{-1} \Breve{\mathbf{A}}_k^H \mathbf{a}_k}
\end{equation}
Therefore, we use the channel powers to develop a user scheduling technique named \textit{channel power-based scheduling} (CPBS).

Let $n \in \integerset_+^*$ be the number of the iteration of the CPBS algorithm. During each iteration, the CPBS schedules the user with the largest channel power, solving the following optimization problem until a stop criterion is met.
\begin{subequations}
\label{eq:problem-cpbs}
\begin{align}
k^* = \; &  \underset{k}{\argmax} \quad P_k\\
& \mathrm{subject\;to} \quad k \in \{1,\dots,K\} \backslash \mathcal{K}^{(n-1)}
\end{align}
\end{subequations}
The pseudocode of the CPBS algorithms is given in Algorithm \ref{alg:cpbs}.
The CPBS algorithm operates with the procedure described in the following. In lines 5 and 6, the agorithm schedules the user with the largest channel power by solving the problem \eqref{eq:problem-cpbs}. The algorithm repeats this procedure until the set of scheduled users results in an infeasible power allocation problem according to the criterion described in Remark 2, or if all the users in the communication cell are scheduled.
Similarly to the CBS algorithm, the set of scheduled users calculated by the CPBS generate an infeasible power allocation problem. Therefore, the procedure derived in Section \ref{sec:obtaining-feasible-set-scheduled-users} must be applied to $\mathcal{K}$ to generate the final set of scheduled users.

\begin{algorithm}[t]
\KwIn{The number of users, $K$, and the channel vectors, $\set{\mathbf{a}_k}_{k=1}^K$}
\KwOut{The set of scheduled users, $\mathcal{K}$}
    
$P_k \gets \norm{\mathbf{a}_k}{2}^2, \; \forall k \in \{1,\dots,K\}$\;
$\mathcal{K}^{(0)} \gets \emptyset$\;

$\Omega^{(0)},n \gets 0$\;
    
\Repeat{$\Omega^{(n+1)} \geq P_{\max}$ or $|\mathcal{K}^{(n)}| = K$}{

$k^* \gets \underset{k}{\argmax} \; P_k, \; k \in \{1,\dots,K\} \backslash \mathcal{K}^{(n)}$\;
$\mathcal{K}^{(n+1)} \gets \mathcal{K}^{(n)} \cup \{ k^* \}$\;
$\Omega^{(n+1)} \gets \Omega^{(n)} + \overline{p}_{k^*}^{\textsc{su}}$\;

$n \gets n + 1$\;
}

$\mathcal{K} \gets \mathcal{K}^{(n)}$\;

\caption{\small \textbf{CPBS} -- Channel power-based scheduling}
\label{alg:cpbs}
\end{algorithm}

\section{Numerical Results}\label{sec:numerical-results}

In this section, we present numerical results to demonstrate the effective performance of the introduced user scheduling methods operating in crowded XL-MIMO systems.
In the Monte-Carlo simulations, we consider $K = 10^3$ users located inside a cell such that $r_k \in [0.03, 1]$ km and $\theta_k \in [-\pi, \pi]$, $\forall k \in \{1,\dots,K\}$.
The users are uniformly distributed in the cell area. Hence, the angles $\theta_k$ follow an uniform distribution, while the distances $r_k$ follow the probability density function \cite{marinello2019}
\begin{equation}
f_{r_k}(r) = \begin{cases}
2r(r_{\max}^2 - r_{\min}^2)^{-1}, & \text{if} \; r_{\min} \leq r \leq r_{\max}\\
0, & \text{otherwise}
\end{cases}
\end{equation}
where $r_{\min}$ and $r_{\max}$ are respectively the minimum and maximum distance from the array center in the communication cell.
The BS is equipped with $M = 10^3$ antennas.
The minimum achievable rate per user is set to $\overline{R}_k \in [5,15]$ bps/Hz, $\forall k \in \{1,\dots,K\}$. We choose the minimum value of 5 bps/Hz in order to meet the ITU-R experienced data rate requirement of 100 Mbps for the dense urban eMBB scenario \cite{itu-m2410}.
On the matter of the channel model, the path-loss attenuation and coefficients are defined according to the ITU-R urban micro-cell environment \cite{itu-m2412}.
The complete list of the simulation parameters is organized in Table \ref{tab:simulation-parameters}.
The evaluation metrics are calculated by averaging the results obtained from $S = 10^3$ realizations. During each realization, the users positions and the channels are generated by sampling random distributions following the definitions provided in Section \ref{sec:system-model}.

\begin{table}[htbp!]
\centering
\caption{Simulation parameters for evaluation of the user scheduling techniques.}
\label{tab:simulation-parameters}
\begin{tabular}{|l|l|}
\hline
\textbf{Parameter} & \textbf{Value}\\
\hline
\multicolumn{2}{|c|}{\textbf{System}}\\
\hline
Number of antennas & $M = 10^3$\\
Carrier frequency & $f_c = 4$ GHz\\
System bandwidth & $B = 20$ MHz\\
Antennas spacing & $d = 3.75$ cm\\
Number of users & $K = 10^3$\\
Minimum achievable rate & $\overline{R}_k {\in [5,15]}$ bps/Hz, {$\forall k$}\\
User distance range & $r_k \in [0.03, 1]$ km\\
User angle range & $\theta_k \in \left[ -\pi, \pi \right]$ m\\
Transmit power budget & $P_{\max} \in {[0, 30]}$ dBm\\
\hline
\multicolumn{2}{|c|}{\textbf{Channel}}\\
\hline
LoS probability & $\rho \in \{0, 0.25, 0.75, 1\}$\\
LoS channel path loss exponent & $\gamma^{\textsc{LoS}} = 2.20$\\
NLoS channel path loss exponent & $\gamma^{\textsc{NLoS}} = 3.67$\\
LoS channel path loss attenuation & $\beta_0^{\textsc{LoS}} = 10^{-4.00}$\\
NLoS channel path loss attenuation & $\beta_0^{\textsc{NLoS}} = 10^{-3.85}$\\
Noise power spectral density & $-174$ dBm/Hz\\
\hline
\multicolumn{2}{|c|}{\textbf{CBS algorithm}}\\
\hline 
Admissibility for channel orthogonality & $\epsilon = 0.4$\\
\hline
\multicolumn{2}{|c|}{\textbf{Monte-Carlo Simulation}}\\
\hline 
Number of realizations & $S = 10^3$\\
\hline
\end{tabular}
\end{table}

\subsection{Evaluation Metrics}\label{sec:evaluation-metrics}

The metrics used to evaluate the user scheduling techniques are \textit{a}) the achievable sum-rate; \textit{b}) the average achievable rate; and \textit{c}) the number of scheduled users. Moreover, we have defined metrics to analyze the \textit{d}) distribution of the scheduled users across the cell, and \textit{e}) the probability of a user being scheduled given its channel state.

From the achievable rate of the user $k$ defined in eq. \eqref{eq:achievable-rate}, the system \textit{achievable sum-rate} is calculated by:
\begin{equation}
\label{eq:achievable-sum-rate}
\mathcal{R} = \sum_{k \in \mathcal{K}} \log_2 \left( 1 + \frac{p_k}{\sigma_w^2 \left[ \left( \mathbf{A}^H \mathbf{A} \right)^{-1} \right]_{k,k}} \right)
\end{equation}
Using eq. \eqref{eq:achievable-sum-rate}, the \textit{average achievable rate} can be expressed dividing the sum-rate by the number of scheduled users:
\begin{equation}
\overline{\mathcal{R}} = \frac{\mathcal{R}}{|\mathcal{K}|}
\end{equation}

To measure the \textit{distribution of the scheduled users} across the cell, we determine the complementary cumulative distribution function (CCDF) of the 2D distance between the scheduled users and the array center. The CCDF for a distance $r \geq 0$ is calculated by
\begin{equation}
\label{eq:ccdf-2d-distances}
\bar{F}(r) = \prob{k \in \mathcal{K} \mid r_k > r}
\end{equation}
From a numerical perspective, the CCDF in eq. \eqref{eq:ccdf-2d-distances} can be approximated by deploying the result of a Monte-Carlo simulation as
\begin{equation}
\widehat{\bar{F}}(r) = \frac{\sum_{\mathcal{K} \in \mathcal{S}} \sum_{k \in \mathcal{K}} I\left( r_k > r \right)}{\sum_{\mathcal{K} \in \mathcal{S}} |\mathcal{K}|}
\end{equation}
where $\mathcal{S} = \set{\mathcal{K}_1,\dots,\mathcal{K}_S}$ is the set containing all the sets of scheduled users obtained in each of $S = |\mathcal{S}|$ Monte-Carlo realizations.

A complementary metric to evaluate the distribution of the scheduled users is the probability of a user being scheduled given its channel state. These probabilities w.r.t. a user under the LoS or NLoS channel state are respectively given by
\begin{align}
P_{\textsc{LoS}} & = \prob{k \in \mathcal{K} \mid x_k = 1}\\
P_{\textsc{NLoS}} & = \prob{k \in \mathcal{K} \mid x_k = 0}
\end{align}
Similarly to eq. \eqref{eq:ccdf-2d-distances}, these two probabilities can be estimated from the result of a Monte-Carlo simulation by calculating:
\begin{align}
\hat{P}_{\textsc{LoS}} & =  \frac{\sum_{\mathcal{K} \in \mathcal{S}} \sum_{k \in \mathcal{K}} I\left( x_k = 1 \right)}{\sum_{\mathcal{K} \in \mathcal{S}} |\mathcal{K}|}\\
\hat{P}_{\textsc{NLoS}} & =  \frac{\sum_{\mathcal{K} \in \mathcal{S}} \sum_{k \in \mathcal{K}} I\left( x_k = 0 \right)}{\sum_{\mathcal{K} \in \mathcal{S}} |\mathcal{K}|}
\end{align}

\subsection{Baseline Techniques}

The baseline user scheduling techniques for XL-MIMO systems adopted for comparison with the proposed CBS and CPBS algorithms include the \textit{greedy weighted clique} (GWC) search algorithm \cite{taesang2005}, the \textit{distance-based scheduling} (DBS), and the \textit{simplified} DBS (s-DBS), both latter proposed in \cite{gonzalez-coma2021}.

In GWC, the users into the communication cell are represented by an UWG. The edges are drawn according to the $\epsilon$-orthogonal rule as described in Section \ref{sec:user-scheduling-clique-search}, while the vertices weights are the single user capacities considering uniform power allocation. The GWC algorithm of \cite{taesang2005} implements a greedy algorithm to search the maximum weighted clique in the graph aiming to obtain a set of scheduled users that have simultaneously high channel powers and quasi-orthogonal channel vectors.

Differently, the DBS algorithm performs user scheduling using a metric named \textit{equivalent distance}, defined in eq. (7) of \cite{gonzalez-coma2021}. The equivalent distance of a given user during an iteration of the DBS algorithm essentially depends on its distance to the center of the BS array and the sum of the inner products between its channel vector and the precoding vectors of the current scheduled users. Hence, users with lower equivalent distance value have higher scheduling priority. The algorithm proceeds selecting the users with the lowest equivalent distance until there is a reduction on the achievable sum-rate.
Alternatively, also in \cite{gonzalez-coma2021} is proposed the s-DBS algorithm, a version of the DBS with lower computational complexity. In this algorithm, the equivalent distance metric is substituted by the distance between the user and the center of the BS array, reducing the complexity of the algorithm at the cost of a performance degradation.

For a fair comparison between the proposed and baseline scheduling techniques, we have included one additional stop criterion on the GWC, DBS, and s-DBS algorithms. During the end of each iteration of the baseline algorithms, the feasibility of the power allocation problem is evaluated by applying eq. \eqref{eq:power-allocation-feasibility}. If this criterion is violated, the last scheduled user is removed from the set and the algorithm stops.
After the user scheduling procedure, the power allocation is carried out by calculating the solution of the optimization problem $\mathcal{P}_2$.

\subsection{User Scheduling Performance}\label{sec:user-scheduling-performance}

\begin{figure}[t]
\centering
\subfigure[Sum-rate]{
\includegraphics[width=.45\columnwidth]{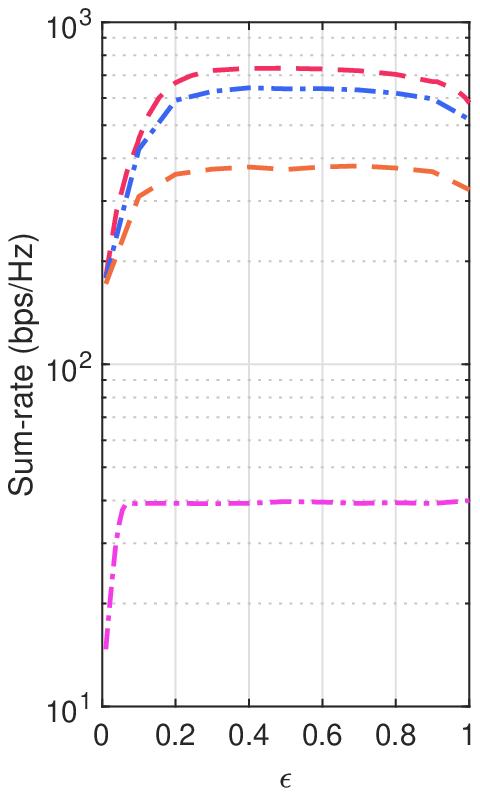}
\label{fig:cbs-epsilon-sum-rate}
}
\subfigure[Number of scheduled users]{
\includegraphics[width=.45\columnwidth]{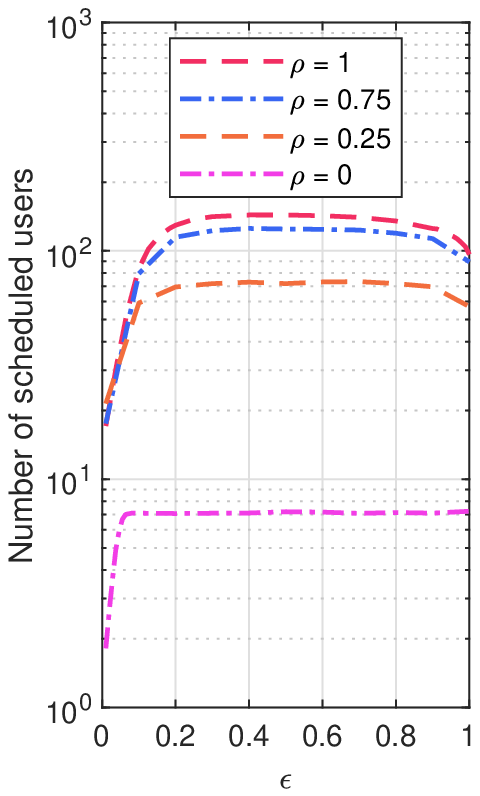}
\label{fig:cbs-epsilon-num-scheduled-users}
}
\caption{Achievable sum-rate and number of scheduled users \textit{vs.} the admissibility channel orthogonality ($\epsilon$) for the proposed CBS algorithm under different LoS probabilities. $P_{\max} = 0$ dBm. $K=10^3$ users; $M= 10^3$ antennas.}
\label{fig:cbs-epsilon}
\end{figure}

Fig. \ref{fig:cbs-epsilon} depicts the achievable sum-rate and the number of scheduled users obtained by the proposed CBS algorithm depending on the parameter of admissibility for channel orthogonality.
Such analysis is paramount to tune the CBS $\epsilon$ parameter for the performance comparison carried out in the following.
From Fig. \ref{fig:cbs-epsilon}, one can see that both the sum-rate and the number of scheduled users are almost constant for $\rho = 0$ and $\epsilon > 0.1$. This occurs because, due to the law of large numbers, the NLoS channel state benefits from the favorable propagation offered by the XL-MIMO array.
Differently, for $\rho > 0$, the peaks of sum-rate and number of scheduled users are achieved in the range $\epsilon \in [0.3, 0.7]$.
In this case, similarly to the result obtained in \cite{taesang2006}, if $\epsilon$ is close to 1, the performance degrades due to the reduction on the effective channel gains paid to obtain the IUI suppression provided by the ZF precoder. On the other hand, if $\epsilon$ is close to 0, the multi-user diversity gain decreases.
Considering this result, we choose $\epsilon = 0.4$ to generate the remaining numerical results.

\begin{figure}[t]
\centering
\subfigure[LoS ($\rho = 1$)]{
\includegraphics[width=.46\columnwidth]{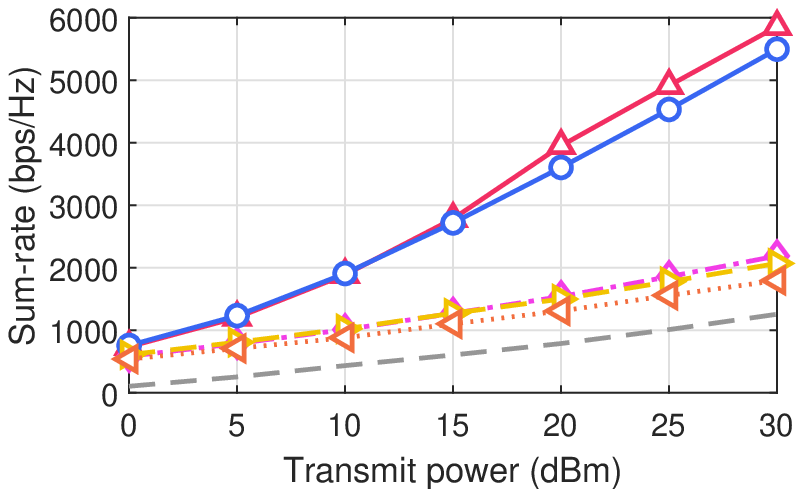}
\label{fig:sum-rate-los}
}
\subfigure[LoS$+$NLoS ($\rho = 0.75$)]{
\includegraphics[width=.46\columnwidth]{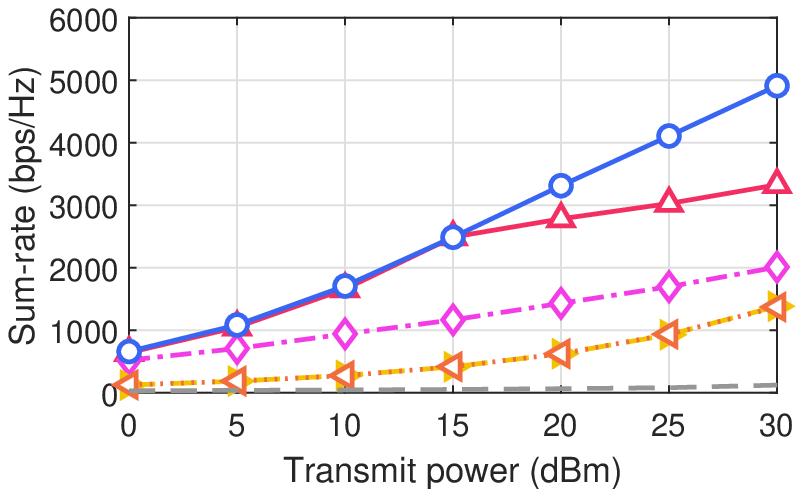}
\label{fig:sum-rate-los-prob-p75}
}
\subfigure[LoS$+$NLoS ($\rho = 0.25$)]{
\includegraphics[width=.46\columnwidth]{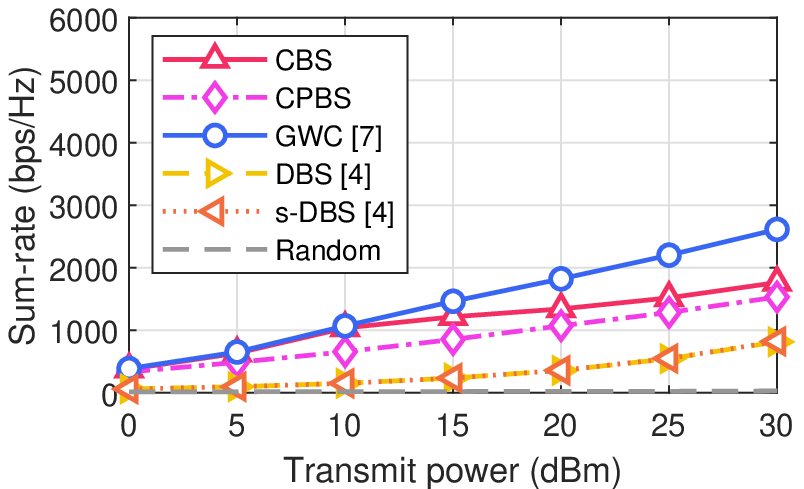}
\label{fig:sum-rate-los-prob-p25}
}
\subfigure[NLoS ($\rho = 0$)]{
\includegraphics[width=.46\columnwidth]{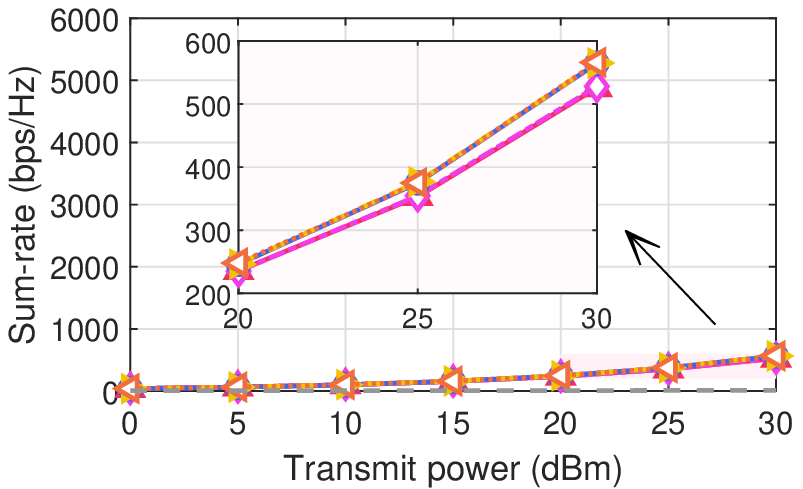}
\label{fig:sum-rate-nlos}
}
\caption{Achievable sum-rate: 
proposed CBS and CPBS algorithms, the GWC algorithm \cite{taesang2005}, the DBS and s-DBS algorithms \cite{gonzalez-coma2021}, and the random user scheduling \textit{vs.} the {transmit power} under different LoS probabilities.}
\label{fig:sum-rate}
\end{figure}

Firstly, we evaluate the user scheduling techniques in terms of the \textbf{achievable sum-rate}. Fig. \ref{fig:sum-rate} depicts the achievable sum-rate obtained by the techniques depending on the {transmit power} and considering different LoS probability values.
At first glance, we see that the LoS probability value changes drastically the performance of the user scheduling algorithms, since the users channel quality depends directly on the channel state. We note that decreasing the LoS probability reduces the achievable sum-rate obtained by all the algorithms. Moreover, as expected, the sum-rate increases with the {transmit power}.
It is worth notice that the graph-based techniques achieve the best sum-rate performance among the evaluated ones. The CBS and GWC algorithms have similar performance for low {transmit power}. However, for $P_{\max} > 15$ dBm and $\rho \leq 0.75$, the GWC algorithm outperforms significantly the CBS one. After the graph-based techniques, the CBS algorithm is the one that achieves the best performance.
The DBS and s-DBS algorithms achieve almost the same performance for all the evaluated cases, outperforming only the random scheduling.
Indeed, it is worth mentioning that, except for the random scheduling, all the evaluated techniques achieve similar performance  for $\rho = 0$.
The random scheduling achieves the poorest performance because its scheduling criterion does not take into account the quality of the users channel vectors. Such behavior repeats for all the numerical results in the sequel.

\begin{figure}[t]
\centering
\subfigure[LoS ($\rho = 1$)]{
\includegraphics[width=.46\columnwidth]{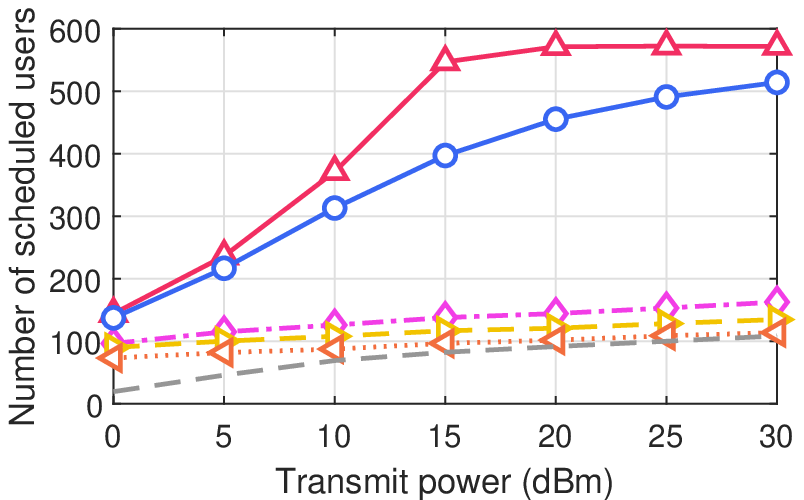}
\label{fig:num-scheduled-users-los}
}
\subfigure[LoS$+$NLoS ($\rho = 0.75$)]{
\includegraphics[width=.46\columnwidth]{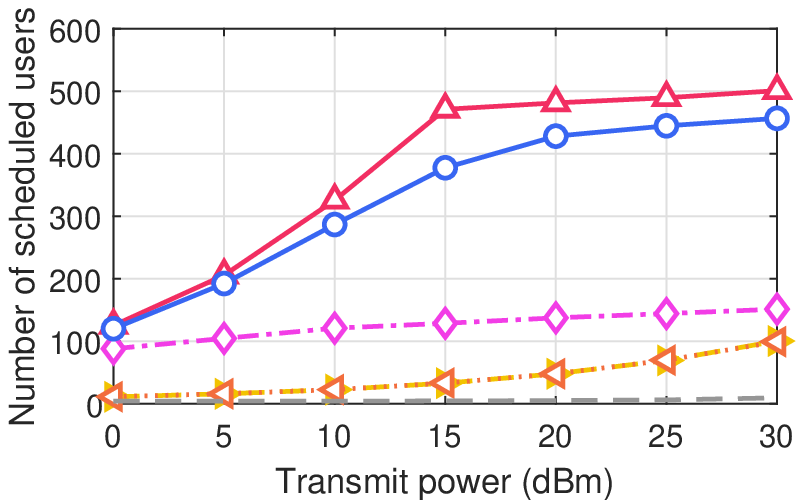}
\label{fig:num-scheduled-users-los-prob-p75}
}
\subfigure[LoS$+$NLoS ($\rho = 0.25$)]{
\includegraphics[width=.46\columnwidth]{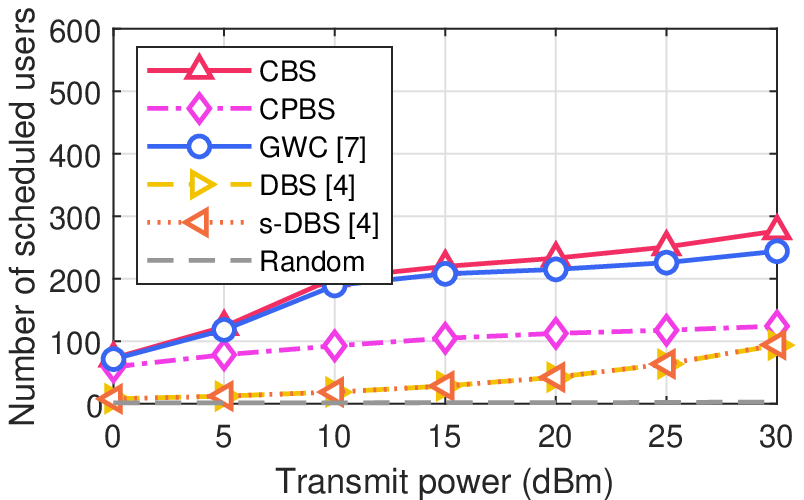}
\label{fig:num-scheduled-users-los-prob-p25}
}
\subfigure[NLoS ($\rho = 0$)]{
\includegraphics[width=.46\columnwidth]{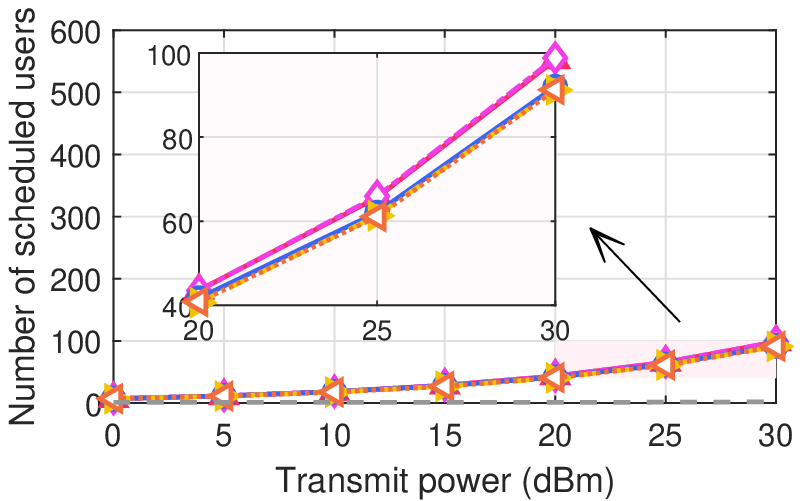}
\label{fig:num-scheduled-users-nlos}
}
\caption{Number of scheduled users: 
proposed CBS and CPBS algorithms, the GWC algorithm \cite{taesang2005}, the DBS and s-DBS algorithms \cite{gonzalez-coma2021}, and the random user scheduling \textit{vs.} the {transmit power} under different LoS probabilities.}
\label{fig:num-scheduled-users}
\end{figure}

Fig. \ref{fig:num-scheduled-users} depicts the \textbf{number of scheduled users} as a function of the {transmit power} considering different values of LoS probability.
One can see that the number of scheduled users increases with the {transmit power} and LoS probability. Specifically for $\rho = 0$, all the evaluated techniques, except for the random scheduling, schedule almost the same number of users.
For $\rho > 0$, the CBS algorithm achieves the best performance in terms of number of scheduled users, followed by GWC. It is worth mentioning that, despite the GWC attains higher sum-rate than the CBS for $P_{\max} > 15$ dBm and $ \rho \in (0;\, 0.75]$, the CBS consistently schedules a higher number of users.
Finally, similarly to what occurs with the sum-rate metric, the number of scheduled users achieved by all the scheduling techniques treated herein, except for the random scheduling, are nearly the same for $\rho = 0$.

\begin{figure}[t]
\centering
\subfigure[LoS ($\rho = 1$)]{
\includegraphics[width=.46\columnwidth]{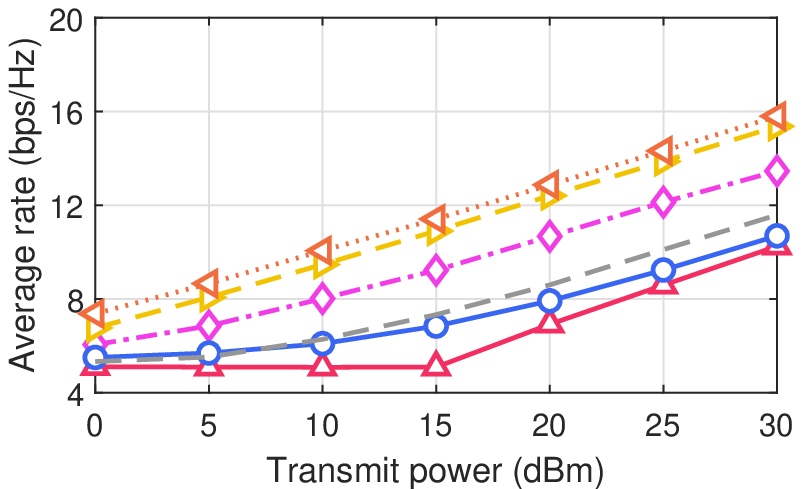}
\label{fig:average-rate-los}
}
\subfigure[LoS$+$NLoS ($\rho = 0.75$)]{
\includegraphics[width=.46\columnwidth]{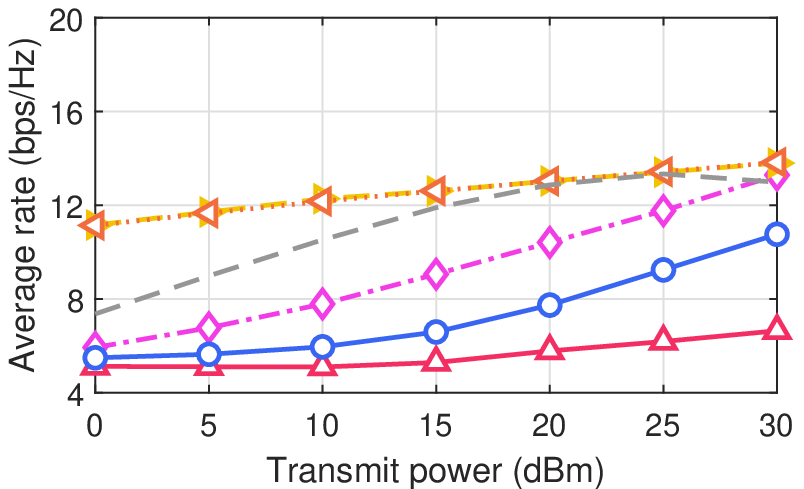}
\label{fig:average-rate-los-prob-p75}
}
\subfigure[LoS$+$NLoS ($\rho = 0.25$)]{
\includegraphics[width=.46\columnwidth]{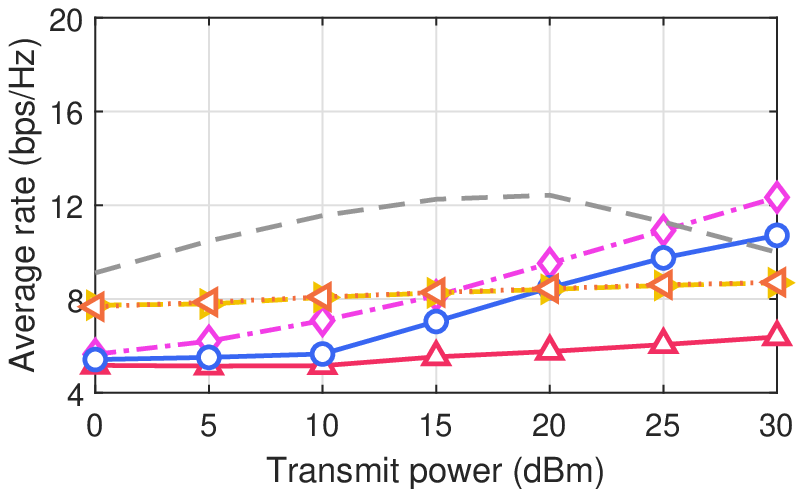}
\label{fig:average-rate-los-prob-p25}
}
\subfigure[NLoS ($\rho = 0$)]{
\includegraphics[width=.46\columnwidth]{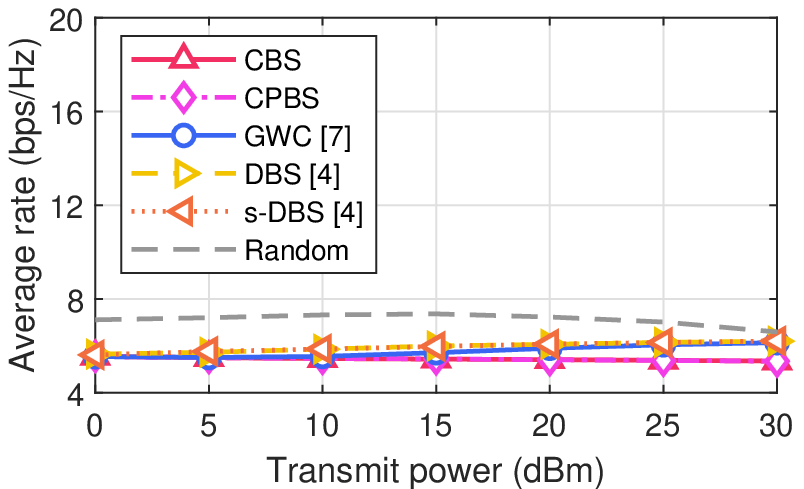}
\label{fig:average-rate-nlos}
}
\caption{Average rate of the scheduled users: proposed CBS and CPBS algorithms, GWC algorithm \cite{taesang2005}, DBS and s-DBS algorithms \cite{gonzalez-coma2021}, and the random scheduling \textit{vs.} the {transmit power} under different LoS probabilities.}
\label{fig:average-rate}
\end{figure}

Fig. \ref{fig:average-rate} depicts the \textbf{average achievable rate} depending on the {transmit power} considering different values of LoS probability.
For all the evaluated techniques, except for the random scheduling, the average rate is inversely proportional to the number of scheduled users. Specifically, for $\rho < 1$ the CBS algorithm obtain achievable rate values near to the minimum achievable rate of 5 bps/Hz.
For $\rho \geq 0.75$, the best techniques in terms of average rate are the DBS and s-DBS. On the other hand, for $\rho \leq 0.25$, the random scheduling achieves the best performance in terms of average rate. However, it is important to mention that this high average rate is obtained at the cost of scheduling an extremely low number of users, as demonstrated in Fig. \ref{fig:num-scheduled-users}.

\begin{figure}[t]
\centering
\subfigure[LoS ($\rho = 1$)]{
\includegraphics[width=.46\columnwidth]{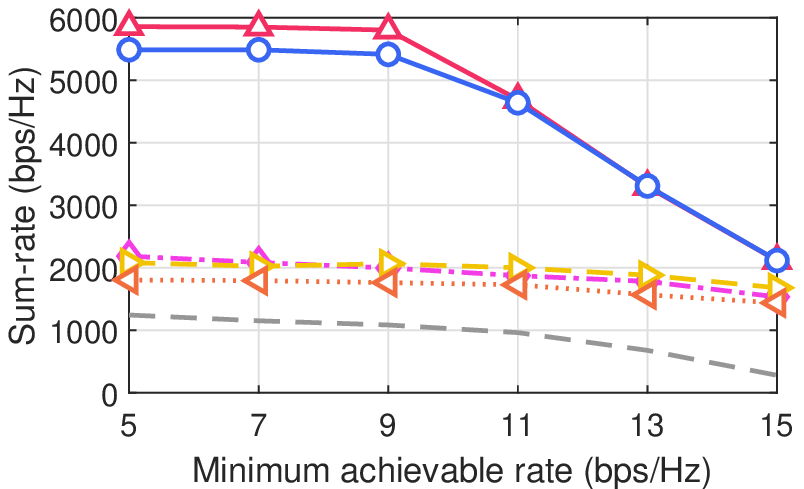}
\label{fig:sum-rate-vs-min-rate-los}
}
\subfigure[LoS$+$NLoS ($\rho = 0.75$)]{
\includegraphics[width=.46\columnwidth]{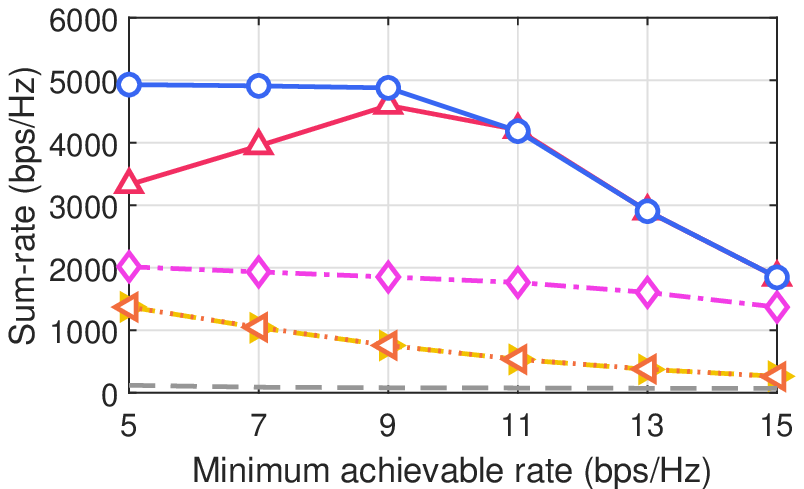}
\label{fig:sum-rate-vs-min-rate-los-prob-p75}
}
\subfigure[LoS$+$NLoS ($\rho = 0.25$)]{
\includegraphics[width=.46\columnwidth]{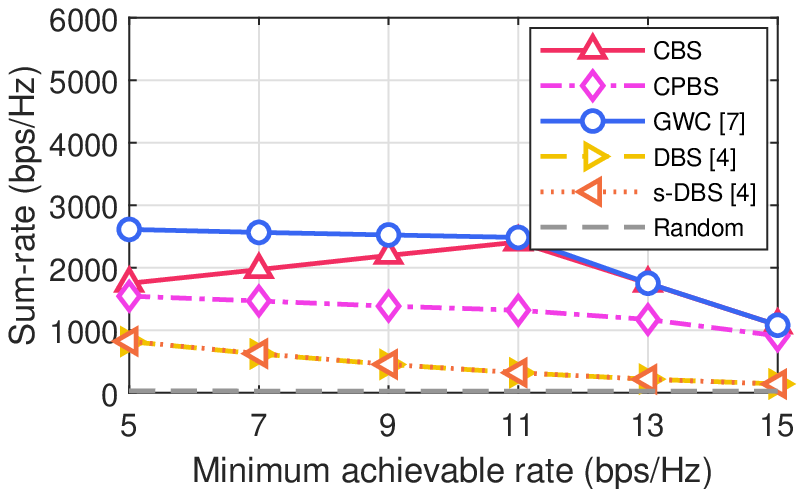}
\label{fig:sum-rate-vs-min-rate-los-prob-p25}
}
\subfigure[NLoS ($\rho = 0$)]{
\includegraphics[width=.46\columnwidth]{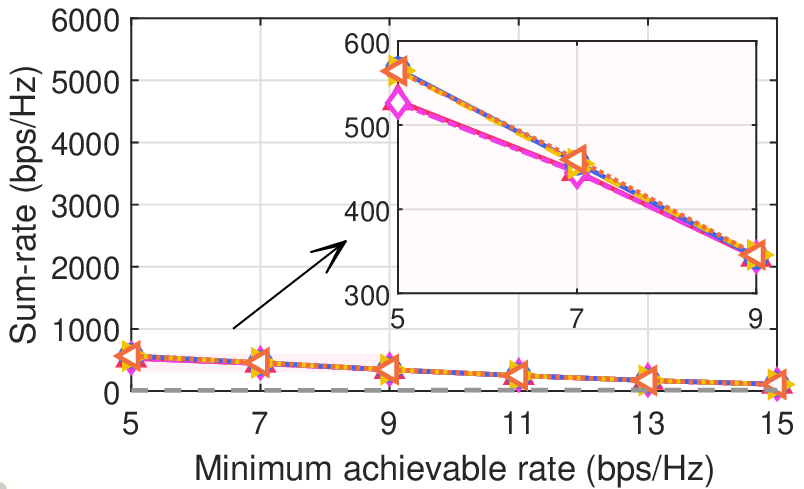}
\label{fig:sum-rate-vs-min-rate-nlos}
}
\caption{Achievable sum-rate \textit{vs.} minimum achievable rate per user under different LoS probabilities. Result of the proposed CBS and CPBS algorithms, the GWC algorithm \cite{taesang2005}, the DBS and s-DBS algorithms \cite{gonzalez-coma2021}, and the random user scheduling. $P_{\max} = 30$ dBm.}
\label{fig:sum-rate-vs-min-rate}
\end{figure}

Fig. \ref{fig:sum-rate-vs-min-rate} depicts the achievable sum-rate as a function of the minimum achievable rate considering different values of LoS probability.
For all the techniques, except for the CBS, the achievable sum-rate decreases by increasing the minimum achievable rate. This behavior is expected, since increasing the minimum achievable rate constraint implies in allocating more power per user to satisfy this requirement.
Specially, the GWC for $\rho \in \{0.25, 0.75\}$ reaches achievable sum-rate values that increase with the minimum achievable rate, up to a point where this behavior reverses. As we will see in the result in the sequel, this maximum point of achievable sum-rate occurs due to a reduction on the number of scheduled users slower than the increase in the minimum achievable rate.

\begin{figure}[t]
\centering
\subfigure[LoS ($\rho = 1$)]{
\includegraphics[width=.46\columnwidth]{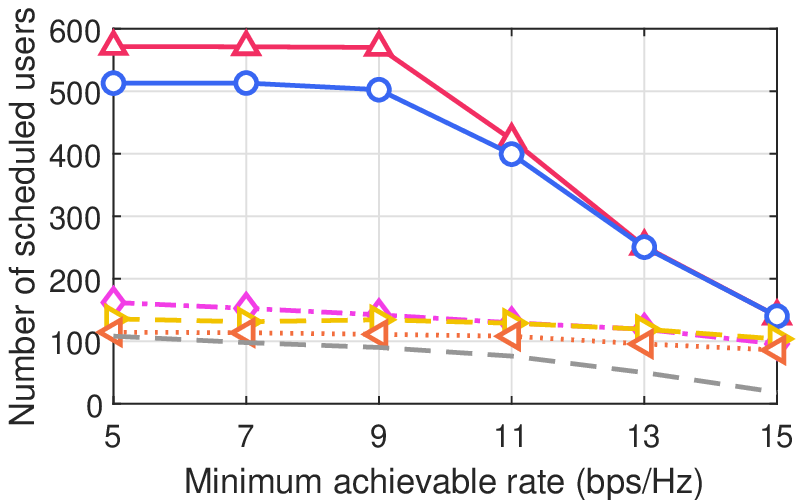}
\label{fig:num-scheduled-users-vs-min-rate-los}
}
\subfigure[LoS$+$NLoS ($\rho = 0.75$)]{
\includegraphics[width=.46\columnwidth]{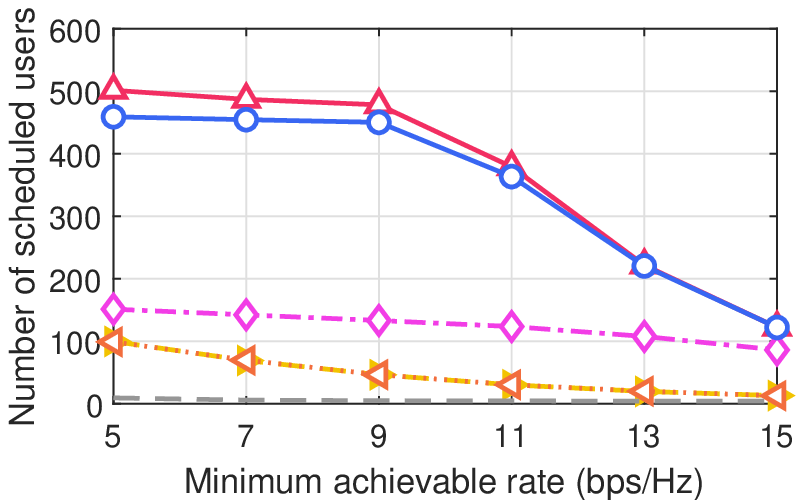}
\label{fig:num-scheduled-users-vs-min-rate-los-prob-p75}
}
\subfigure[LoS$+$NLoS ($\rho = 0.25$)]{
\includegraphics[width=.46\columnwidth]{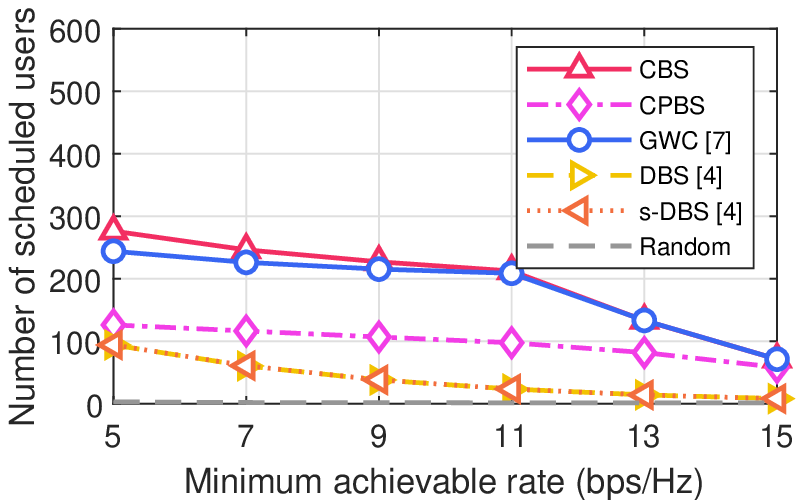}
\label{fig:num-scheduled-users-vs-min-rate-los-prob-p25}
}
\subfigure[NLoS ($\rho = 0$)]{
\includegraphics[width=.46\columnwidth]{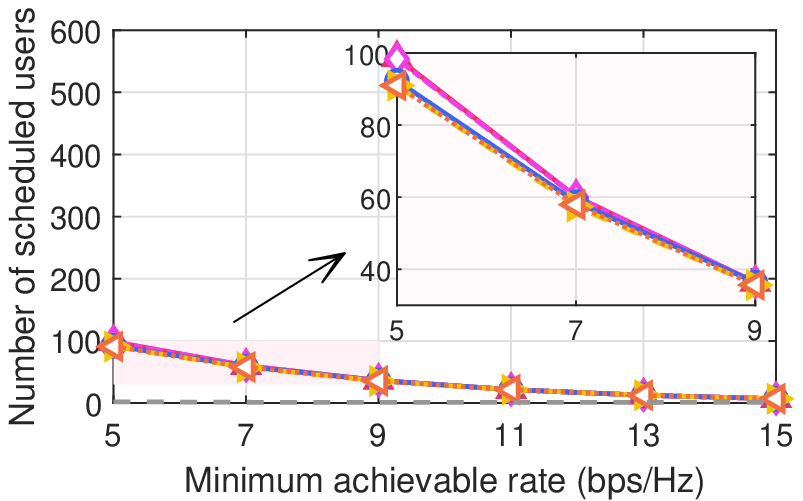}
\label{fig:num-scheduled-users-vs-min-rate-nlos}
}
\caption{Number of scheduled users \textit{vs.} minimum achievable rate under different LoS probabilities. Result of the proposed CBS and CPBS algorithms, the GWC algorithm \cite{taesang2005}, the DBS and s-DBS algorithms \cite{gonzalez-coma2021}, and the random user scheduling. $P_{\max} = 30$ dBm.}
\label{fig:num-scheduled-users-vs-min-rate}
\end{figure}

Fig. \ref{fig:num-scheduled-users-vs-min-rate} depicts the number of scheduled users depending on the minimum achievable rate considering different values of LoS probability.
As expected, the stricter minimum achievable rate constraints with a fixed transmit power budget results in a reduction on the number of scheduled users.
Specifically, we see that the CBS and GWC algorithms present slow rate of decrease in the number of scheduled users for $\rho = 0.75$ and $\overline{R}_k \leq 9$ bps/Hz, $\forall k$, and for $\rho = 0.25$ and $\overline{R}_k \leq 11$ bps/Hz, $\forall k$. This is the cause of the partially increasing behavior of the achievable sum-rate obtained by these algorithms identified in Figs. \ref{fig:sum-rate-vs-min-rate-los-prob-p75} and \ref{fig:sum-rate-vs-min-rate-los-prob-p25}.

\subsection{Distribution of the Scheduled Users}

\begin{figure}[t]
\centering
\subfigure[LoS ($\rho = 1$)]{
\includegraphics[width=.46\columnwidth]{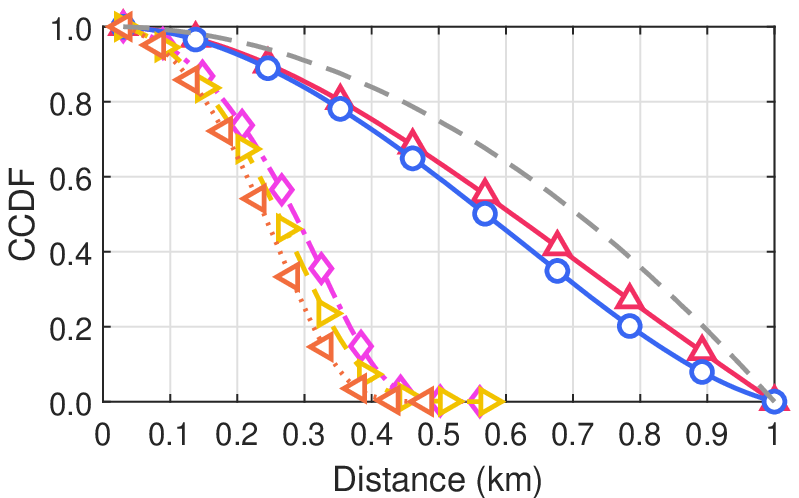}
\label{fig:distribution-scheduled-users-los}
}
\subfigure[LoS+NLoS ($\rho = 0.75$)]{
\includegraphics[width=.46\columnwidth]{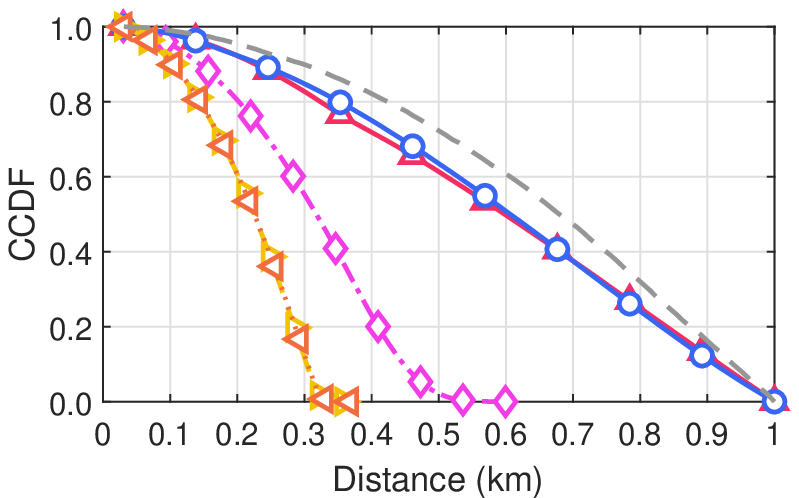}
\label{fig:distribution-scheduled-users-los-prob-p75}
}
\subfigure[LoS+NLoS ($\rho = 0.25$)]{
\includegraphics[width=.46\columnwidth]{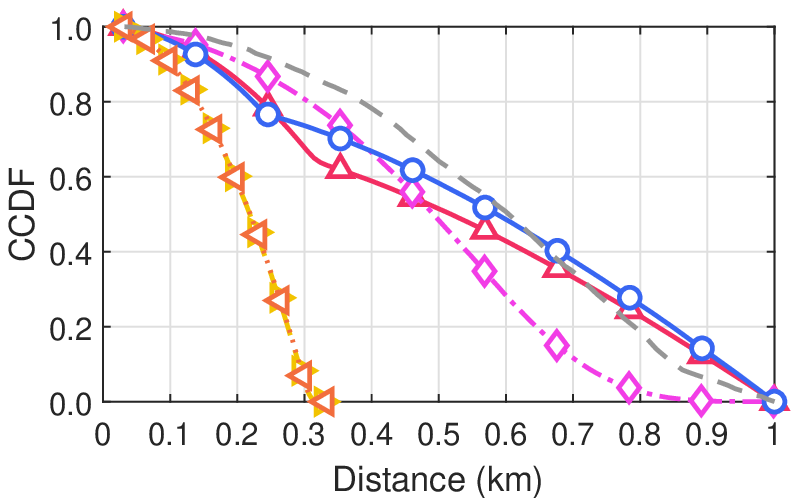}
\label{fig:distribution-scheduled-users-los-prob-p25}
}
\subfigure[NLoS ($\rho = 0$)]{
\includegraphics[width=.46\columnwidth]{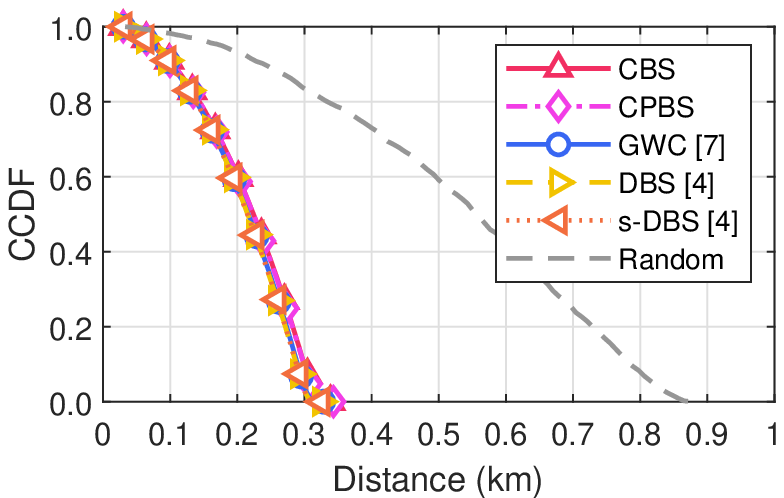}
\label{fig:distribution-scheduled-users-nlos}
}
\caption{CCDF of the 2D distance between the scheduled users and the array center for the proposed CBS and CPBS algorithms, the GWC algorithm \cite{taesang2005}, the DBS and s-DBS algorithms \cite{gonzalez-coma2021}, and the random user scheduling under different LoS probabilities. $P_{\max} = 30$ dBm.}
\label{fig:distribution-scheduled-users}
\end{figure}

Fig. \ref{fig:distribution-scheduled-users} depicts the CCDF of the 2D distance between the scheduled users and the array center considering different values of LoS probability.
For $\rho = 1$, both graph-based user scheduling techniques, our proposed CBS and the GWC \cite{taesang2005}, provide a much higher coverage when compared with the DBS, s-DBS, and CPBS algorithms, allowing scheduling users located at cell border ($0.8 < r_k\leq 1.0$ km). In fact, the latter algorithms schedule only users that are around 0.5 km apart from the array, half the cell radius. This demonstrate the superiority of the proposed CBS algorithm in providing a more uniform communication experience for users positioned throughout the communication cell, including border users. On the other hand, for $\rho = 0$, one can see that, except for the random scheduling, all the evaluated techniques achieve the same poor coverage, scheduling users only around 0.3 km apart from the array center. Specifically, this result occurs due to the high path-loss associated with the NLoS channel state and the limited {transmit power}.
Particularly, we note that the random scheduling attains the best coverage in all the evaluated LoS channel probability cases. However, such good coverage comes at the price of low performance in terms of achievable sum-rate and number of scheduled users, as demonstrated in Figs. \ref{fig:sum-rate} and \ref{fig:num-scheduled-users}.

\begin{figure}[t]
\subfigure[LoS+NLoS ($\rho = 0.75$)]{
\includegraphics[width=.46\columnwidth]{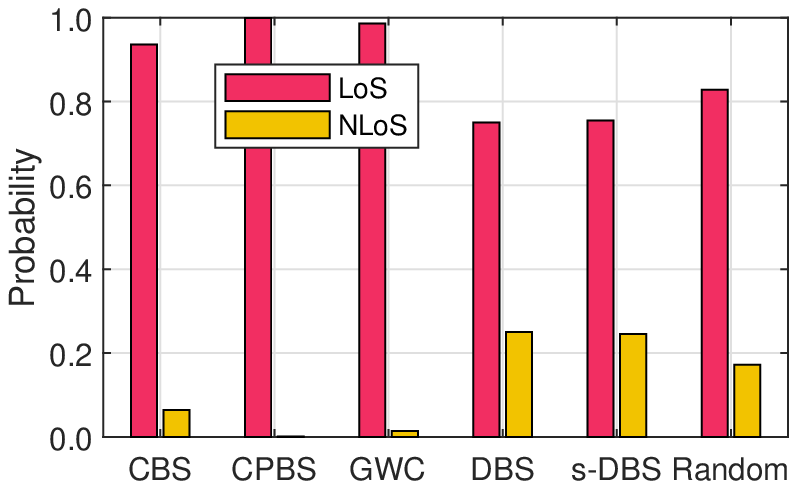}
\label{fig:distribution-scheduled-users-channel-state-los-prob-p75}
}
\subfigure[LoS+NLoS ($\rho = 0.25$)]{
\includegraphics[width=.46\columnwidth]{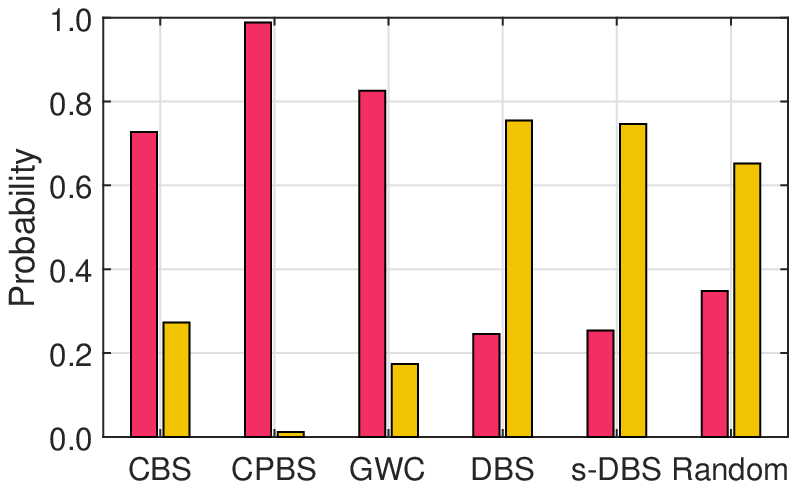}
\label{fig:distribution-scheduled-users-channel-state-los-prob-p25}
}
\caption{Probability of a user being scheduled given its channel state for the proposed CBS and CPBS algorithms, the GWC algorithm \cite{taesang2005}, the DBS and s-DBS algorithms \cite{gonzalez-coma2021}, and the random user scheduling under different LoS probabilities.}
\label{fig:distribution-scheduled-users-channel-state}
\end{figure}

Fig. \ref{fig:distribution-scheduled-users-channel-state} depicts the probability of scheduling users under LoS and NLoS channel states considering different values of LoS probability.
The techniques that consider the minimum required power to attain the minimum QoS and the channel quality tend to schedule users in the most favorable channel state, namely the LoS state. On the other hand, the techniques that take the distance into consideration tend to schedule users in the same proportion as that they appear in the communication cell.

From the numerical results presented in Figs. \ref{fig:sum-rate}--\ref{fig:distribution-scheduled-users-channel-state} one can conclude that the CBS algorithm can achieve high numbers of scheduled users and provide fair coverage when users under the both LoS and NLoS channel states coexist in the same communication cell, \textit{i.e.}, $\rho > 0$.
On the matter of the CPBS, it outperforms the DBS and s-DBS algorithms in all the evaluated scenarios, except for $\rho = 0$. However, the scheduling prioritization based on the channel power of CPBS significantly reduces the coverage.
For this reason, in the context of XL-MIMO systems operating in crowded communications scenarios, the CBS algorithm is a promising technique for scheduling high numbers of users satisfying the minimum QoS constraints, while providing fair coverage over the whole cell area.

\section{Conclusions}\label{sec:conclusions}

In this work, we propose a channel model for XL-MIMO arrays based on the {SW} propagation model considering that users under LoS and NLoS channel states coexist in the same communication cell.
Indeed, we evaluate the performance of user-scheduling techniques under a broad range of channel state conditions, by characterizing the distribution of the scheduled users inside the cell.
The numerical results on the achievable sum-rate, average rate of the scheduled users, the distribution of the scheduled users across the
cell, and number of scheduled users demonstrate that the user-scheduling performance is sensible to the probability of a user having channel in the LoS state.
Interesting, our proposed CBS method depending on the parameter of admissibility for channel orthogonality, also revealing a clear superiority in terms of number of scheduled users under minimum user-rate and transmit power budget constraints. Moreover, the proposed CBS method is able to schedule border user satisfactory providing a more homogeneous service for both center and border users.

\end{document}